\documentclass[]{aa}  

\usepackage{graphicx}
\usepackage{txfonts}
\newcommand{\emma}{{\texttt{EMMA} }}
\authorrunning{Deparis et al.}
\begin{document}

   \title{Impact of the reduced speed of light approximation on ionization front velocities in cosmological simulations of the epoch of reionization}
\titlerunning{Ionization front speeds in reionization simulations}

   \author{Nicolas Deparis,
          \inst{1}
          Dominique Aubert\inst{1},
          Pierre Ocvirk\inst{1},
          Jonathan Chardin\inst{1}
            \and
            Joseph Lewis\inst{1}
          }

   \institute{ Observatoire Astronomique de Strasbourg, CNRS UMR 7550, Universite de Strasbourg, Strasbourg,France\\
             }

   \date{Received, accepted}

 
  \abstract
   {Coupled radiative-hydrodynamics simulations of the epoch of reionization aim to reproduce the propagation of ionization fronts during the transition before the overlap of HII regions. Many of these simulations use moment-based methods to track radiative transfer processes using explicit solvers and are therefore  subject to strict stability conditions regarding the speed of light, which implies a great computational cost. The cost can be reduced by assuming a reduced speed of light, and this approximation is now widely used to produce large-scale simulations of  reionization.}
   {We measure how ionization fronts propagate in simulations of the epoch of reionization. In particular, {we want} to distinguish between the different stages of the fronts' progression into the intergalactic medium.  We also investigate how these stages and their properties are impacted by the choice of a reduced speed of light.}
   {We introduce a new method for estimating and comparing the ionization front speeds based on maps of the reionization redshifts. We applied it to a set of cosmological simulations of the reionization using a set of reduced speeds of light, and measured the evolution of the ionization front speeds during the reionization process. We only considered models where the reionization is driven by the sources created within the simulations, without potential contributions of an external homogeneous ionizing background.}
   {We find that ionization fronts progress via a two-stage process, the first stage at low velocity as the fronts emerge from high density regions and a second later stage just before the overlap, during which front speeds increase close to the speed of light. {For example, using a set of small $8\mathrm{Mpc/h}^3$ simulations}, we find that a minimal velocity of $0.3c$ is able to model these two stages {in this specific context} without significant impact. Values as low as $0.05c$ can model the first low velocity stage, but limit the acceleration at later times. Lower values modify the distribution of front speeds at all times. {Using another set of simulations with larger $64\mathrm{Mpc/h}^3$ volumes {that better account for distant sources}, we find that reduced speed of light has a greater impact on reionization times and front speeds in underdense regions that are reionized at late times and swept by radiation produced by distant sources. Conversely, the same quantities measured in dense regions with slow fronts are less sensitive to $\tilde c$ values. {While the discrepancies introduced by reduced speed of light could be reduced by the inclusion of an additional UV background, we expect these conclusions to be robust in the case of simulations with reionizations driven by inner sources.}}}
   {}

   \keywords{Cosmology: dark ages, reionization, first stars- methods: numerical}

   \maketitle
%

\section{Introduction}
The reionization of the Universe is driven by the propagation of ionization fronts created by the first astrophysical light sources. This process is not homogeneous as it reflects the distribution of absorbers and sources created by the rise of large-scale structures: the typical associated picture consists of a network of HII regions that eventually overlap at the end of reionization, by $z\sim6$.

As a consequence, great efforts are currently being made to model this complex process using cosmological simulations by including radiative transfer physics. Many of these works rely on a moment-based description of radiative transfer (see, e.g., \citet{bauer_hydrogen_2015}, \citet{gnedin_cosmic_2014}, \citet{2016MNRAS.463.1462O}, \citet{2018arXiv180107259R}, \citet{2018arXiv180201613A}), which is described as a set of conservative equations on radiative quantities (e.g., radiative density, radiative flux, radiative pressure). In many aspects, this technique is akin to  modeling  a ``photon fluid''.

A practical advantage of this description is the possibility to use ready-made generic methods or existing modules developed for hydrodynamics. However, moment-based methods have to assume a finite speed of light  as they model the actual propagation of radiation density. Since this fluid description is often handled through an explicit approach, this choice of methodology is not without consequences (even though implicit solvers do exist; see, e.g., \citet{2007A&A...464..429G}). The main constraint for explicit moment-based solvers is related to the Courant-Friedrichs-Lewy (CFL) condition. It imposes that the higher the velocity of a process is, the higher the temporal resolution has to be to guarantee numerical stability (see, e.g., \citet{aubert_radiative_2008,rosdahl_ramses-rt:_2013}).
The time step $\Delta t $ must satisfy
\begin{equation}
\Delta t =  C_{\mathrm{CFL}} \cdot \frac{\Delta x}{v},
\end{equation}
where $C_{\mathrm{CFL}}<1$ (in 1D), $\Delta x$ is the spatial resolution, and $v$ is the maximum speed of the considered process. 
For high velocities, and in the most extreme case for light speed velocities, such solvers are conditionally stable only if the time resolution is consequent, leading to a large number of time steps and a great computational cost.

Several methodologies have emerged to reduce this cost. The first  consists in taking advantage of the recent evolution in hardware; for example, using graphical processing units (GPUs) can divide the computational time by almost two orders of magnitude due to their highly parallel capability \citep{2010ApJ...724..244A}.
The second method is the subject of this study: the reduced speed of light approximation (RSLA).
The RSLA considers that  light propagates at a fraction of its actual speed
\begin{equation}
v=\tilde{c} \cdot c,
\end{equation}
where $c$ is the physical speed of light (299 792 458 m/s) and $\tilde{c} \leq 1$ is the RSLA factor.
The main idea of this technique relies on the fact that the typical velocities of dynamical processes (between $100$ and $10^4$ km/s) can be several orders of magnitude lower than the physical speed of light. As the speed of light is significantly greater than any other speed, lowering its value in a simulation should not have a significant impact, but should allow a comfortable gain on the computational cost.  For example, if the reionization properties remain similar while using $\tilde{c}=1$ or $\tilde{c}=0.1$,  a factor of ten in the computation of radiative processes can be gained just by reducing the number of radiative time steps. This gain can be  appreciable as radiation represents approximately $80 \%$ of the computational cost in this kind of simulation when $\tilde{c}=1$.  {This is the approach used by  \citet{2017MNRAS.468.4831K}, among others, where different values of $\tilde c$ are used in different regimes of density within a given experiment, thus capturing typical front speeds in the relevant regimes while remaining computationally reasonable (see also \citet{2018MNRAS.479..994R}).}  However, the consequences of such an approximation and their dependencies (e.g., on implementation, resolution, or box size) are still debated (\citet{gnedin_multi-dimensional_2001}, \citet{gnedin_proper_2016},\citet{bauer_hydrogen_2015}, \citet{2018MNRAS.479..994R}, Ocvirk et al.  in prep.).

In this study, we  investigate the validity of this approximation and we explore how reducing the speed of light influences the evolution of ionization fronts during the reionization epoch. First, we present a set of radiative hydrodynamics (RHD) simulations with different speeds of light,  and introduce a method for computing the ionization front (hereafter I-front) speeds using reionization redshift maps. Then, using different diagnostics (average ionization history, ionization maps, front speeds), we quantify the impact of RSLA on the global reionization. 
\section{Methodology}

Simulations were produced with \emma {(Electromagnetisme et M\'ecanique sur Maille Adaptative) an adaptive mesh refinement (AMR)} cosmological code with fully coupled RHD \citep{aubert_emma:_2015}.
The light was considered  a fluid; its propagation was tracked using the moment-based M1 approximation \citep{1984JQSRT..31..149L,2007A&A...464..429G,aubert_radiative_2008}. We ran a set of small simulations at a similar resolution of production runs such as  CODA \citep{2016MNRAS.463.1462O} or CROC \citep{gnedin_cosmic_2014}.
We  considered $\left(8h^{-1} \mathrm{cMpc}\sim 12 \mathrm{cMpc}    \right)^3$ volumes, simulated from redshift z=150 to the end of the reionization ($5<z<6$).
Dark matter was resolved with $256^3$ particles with a mass of $3.4 \cdot 10^6 M_\odot$.
Hydrodynamics and radiation were simulated on a grid of $256^3$ resolution elements for a coarsest spatial resolution of 46 ckpc.
The grid was refined according to a semi-Lagrangian scheme when a cell contained a mass greater than eight dark matter particles;  the refinement was not allowed if the spatial resolution of the newly formed cells was lower than 500 pc.
Our stellar mass resolution was set to be equal to $7.2 \cdot 10^4 M_\odot$. Our emissivity model follows Starburst99, with a population of $10^6 M_\odot$ having a Top-Heavy initial mass function and a Z=0.001 metallicity. 
To limit the number of unknowns, we did not consider supernovae feedback  and the corresponding module was turned off. All simulations were run with the exact same parameters (including star formation) and initial conditions except for $\tilde{c}$, the ratio of the simulated speed of light to the real speed of light. 
{The boundary conditions were assumed to be periodic for all processes, including radiative transfer. Radiation from source outside the volume was therefore modeled implicitly by exact replicates of the inner distribution of sources. We assess the impact of a less trivial contribution of  external sources  in section \ref{s:sim64}.} {We did not consider the influence of an additional homogeneous background.

We ran a set of six simulations with $\tilde{c}=1$, $\tilde{c}=0.3$, $\tilde{c}=0.1$, $\tilde{c}=0.05$, $\tilde{c}=0.02$, and $\tilde{c}=0.01$. The $\tilde c$ value was used to solve the set of conservative equations of the M1 approximation (on radiative densities and fluxes) in all regimes and in the calculations of photoionization rates;   we followed the same methodology as \cite{rosdahl_ramses-rt:_2013} and \cite{bauer_hydrogen_2015}.  We note that this methodology differs from that of \citet{gnedin_proper_2016}, who distinguishes between the radiation background and the fluctuations {superimposed}. In this case $\tilde{c}$ is only used to propagate the fluctuations.


\section{Results}

\subsection{Cosmic star formation and ionization histories}

\begin{figure}
    \centering
    \includegraphics[width=0.45\textwidth]{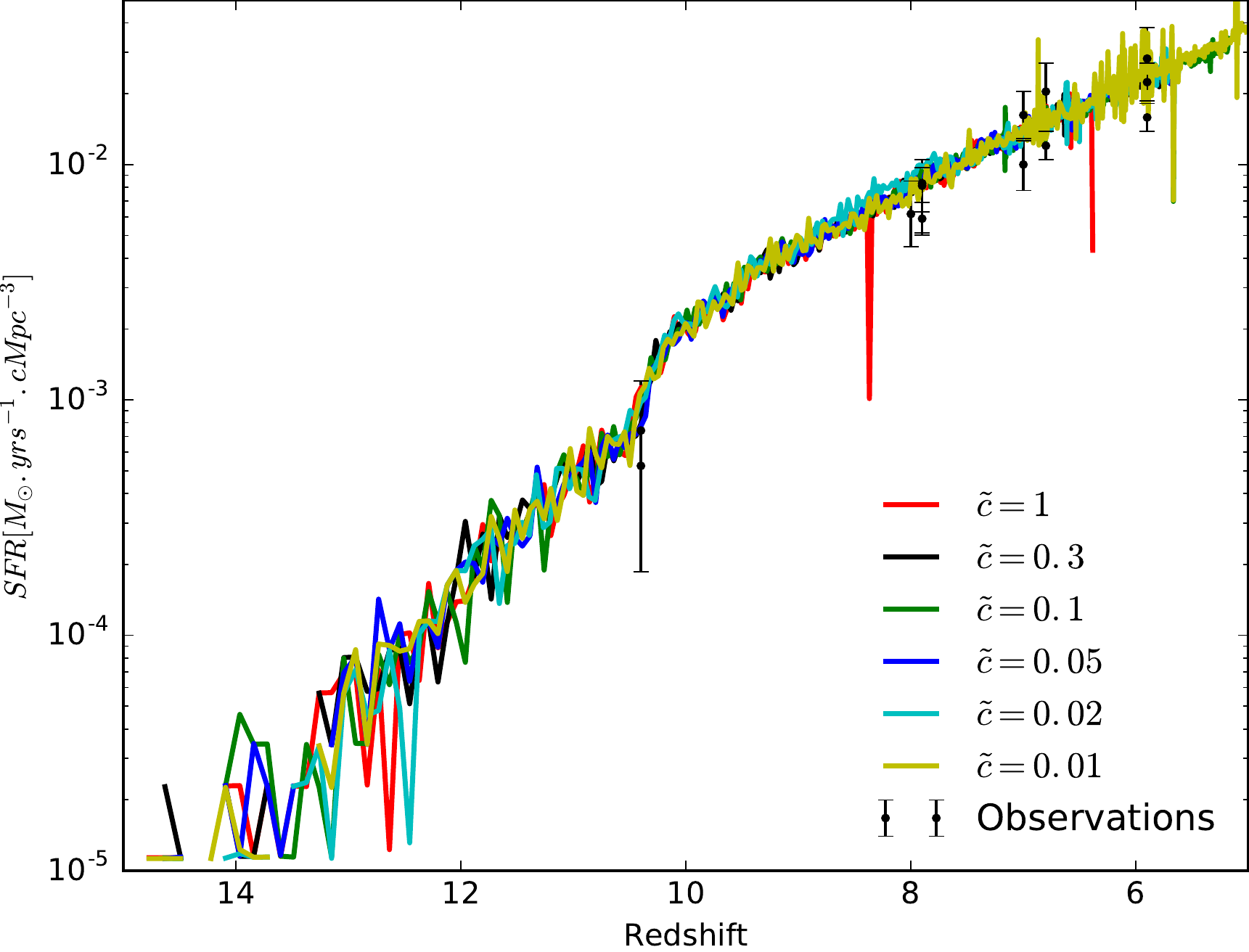} 
    \includegraphics[width=0.45\textwidth]{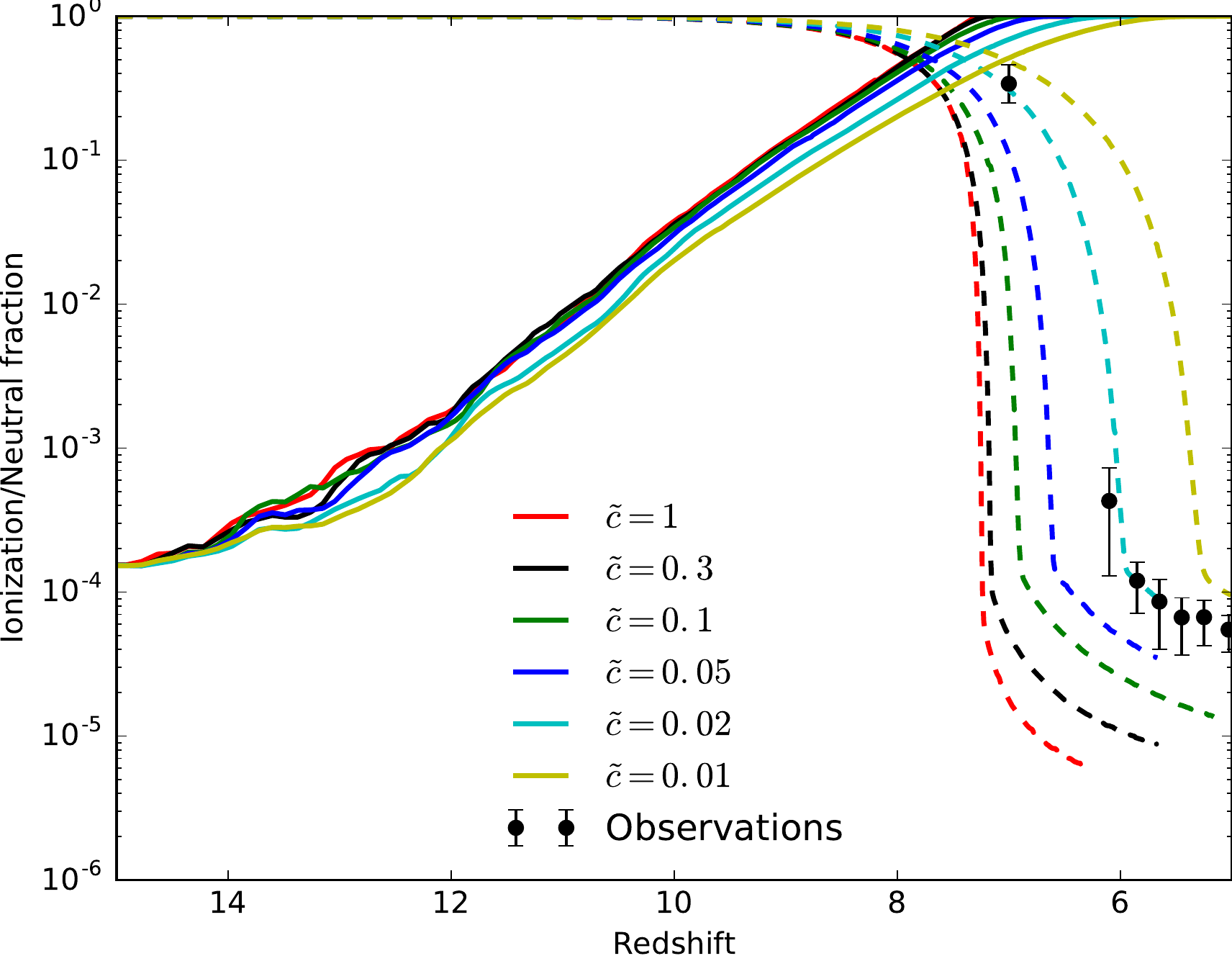}
    \includegraphics[width=0.48\textwidth]{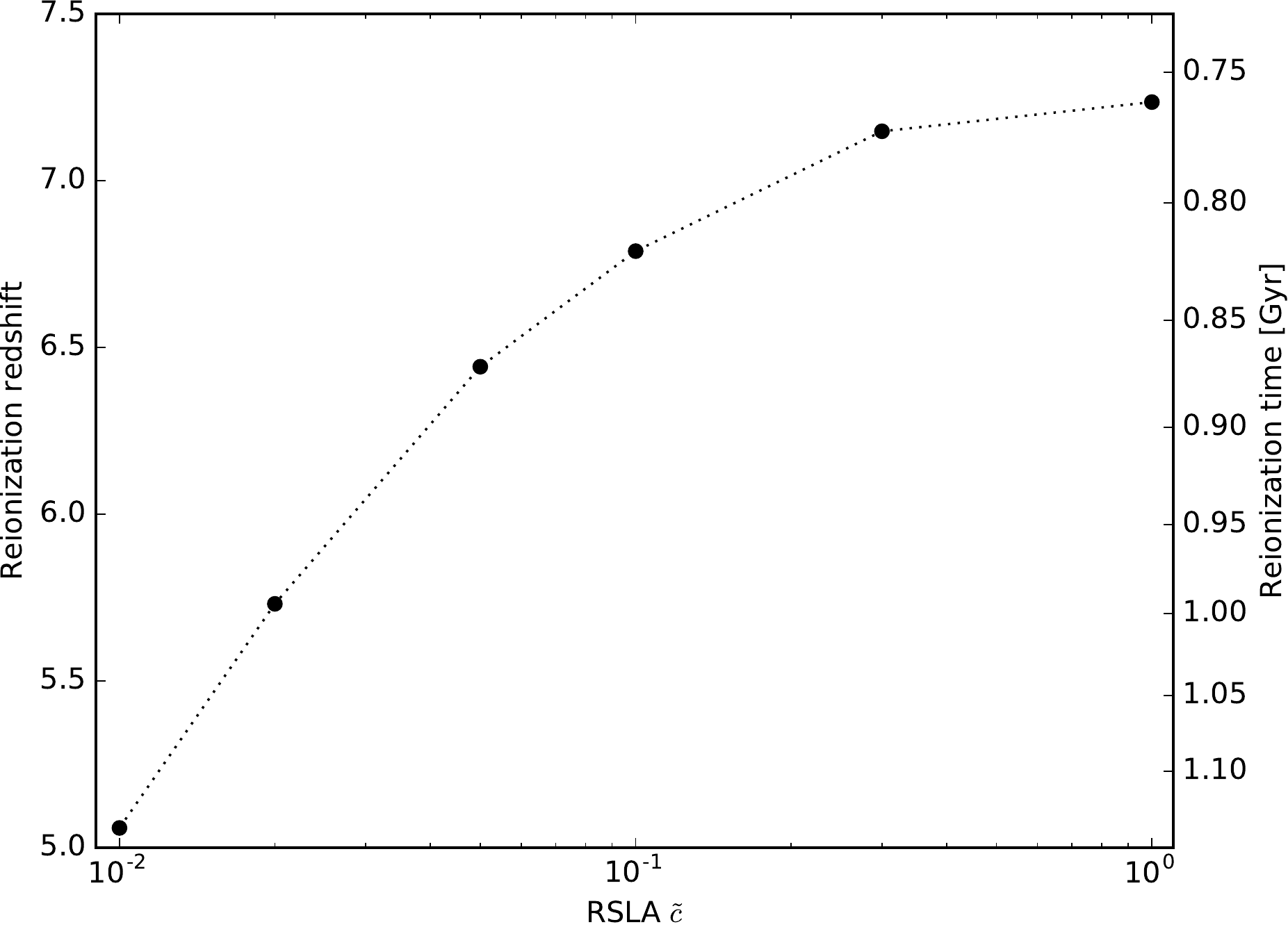}
    \caption{ (Top) Cosmic star formation histories with observational constraints from \citet{bouwens_uv_2015}; (middle) volume weighted ionization fraction (solid lines) and neutral fraction (dashed lines) with observational constraints from \citet{fan_observational_2006} as a function of redshift for different reduced speeds of light;
        (bottom) $x_\mathrm{HII}=0.01\%$ reionization redshifts measured for different reduced speeds of light.
    \label{img_SFR}
    }
    \label{fig:SFRX}
\end{figure}

Figure \ref{fig:SFRX} presents the cosmic star formation histories (SFH) for different $\tilde c$ values.
This measure shows that the RSLA does not change the cosmic SFH. As ionizing photons are emitted by newly formed stars, the photons budget is not impacted by the RSLA and all the simulations have to propagate, at least on average, the same quantity of radiative energy. Figure \ref{fig:SFRX} also presents the average volume weighted ionization state as a function of time $X_V (t)$ 

\begin{equation}
X_V (t)= \frac{ \int x_{(t)} \cdot dV }{ \int dV },
\end{equation}with the local ionization fraction 
\begin{equation}
x_{(t)} =   \frac {n_{H+}}{n_H}
\end{equation}
We observe a direct link between the ionization history and the RSLA.
The slower the light is, the later the reionization occurs.
We define the reionization redshift as the redshift when the volume weighted hydrogen neutral fraction decreases below $X_V=10^{-4}$. At these times the percolation process of HII regions (also known as {the overlap}) is complete.
Figure \ref{fig:SFRX} presents this reionization redshift as a function of the reduced value of the  speed of light.
Reionization redshifts converge as $\tilde{c}$ is close to unity and the slope flattening indicates that any $\tilde{c}>0.1$ leads to  similar instants of reionization.
The other way round, reionization redshifts rapidly decline with decreasing $\tilde{c}$ values.
For instance, $\tilde{c}=0.1$ presents a delay of $\approx 60$ Myr and less than $0.5$ in redshift, but these values are extended to $\approx 425$ Myr and almost 2.2 in redshift for the  $\tilde{c}=0.01$ run. The different reionization histories obtained here suggest that I-fronts may  at some point become as fast as the assumed speed of light, and the RSLA may therefore have an impact on simulation predictions. We assess this point directly in the next sections.


\subsection{Ionization maps}

\begin{figure}
    \center
        \includegraphics[width=0.4\textwidth]{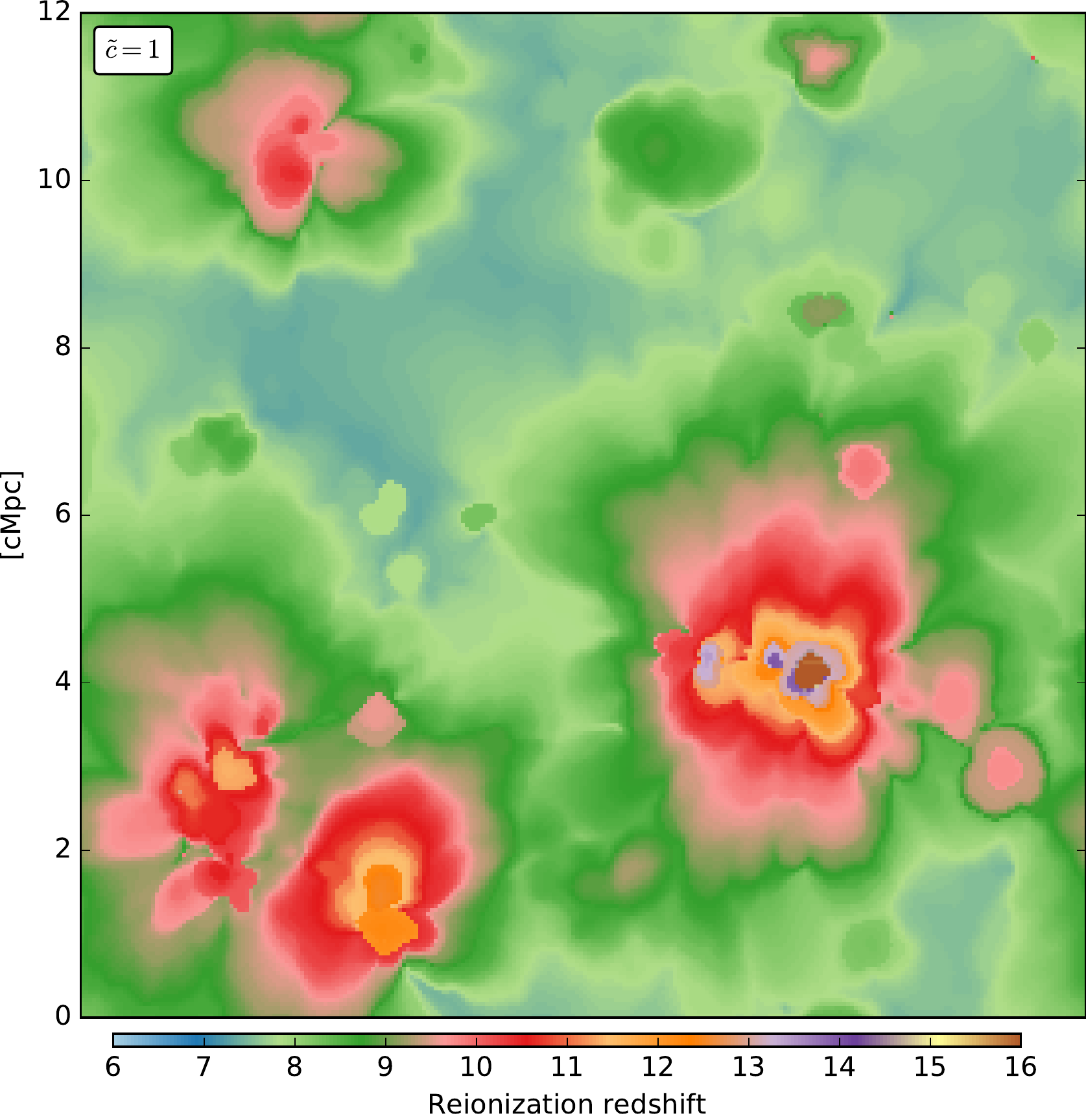}
        \includegraphics[width=0.4\textwidth]{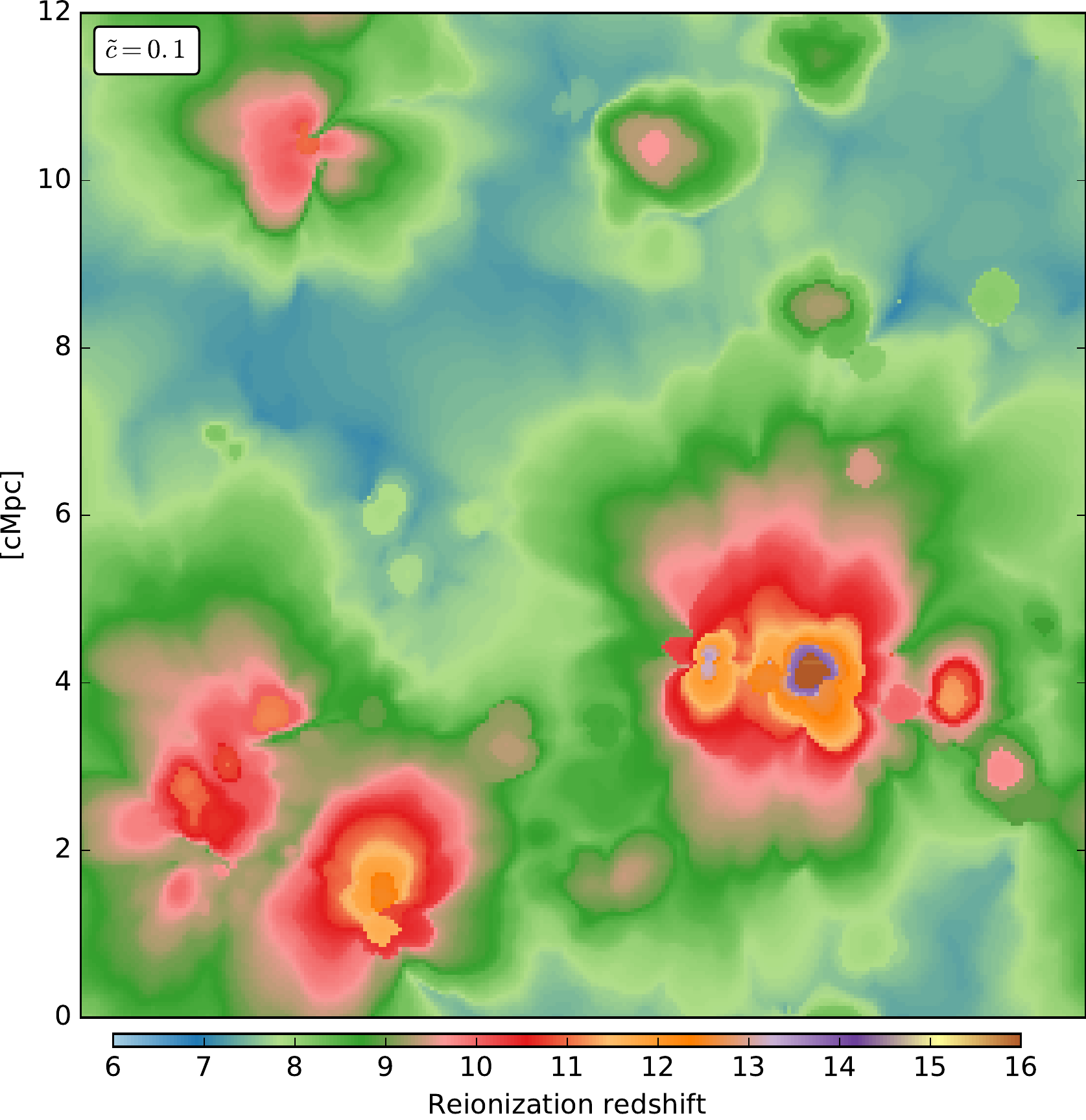}
        \includegraphics[width=0.4\textwidth]{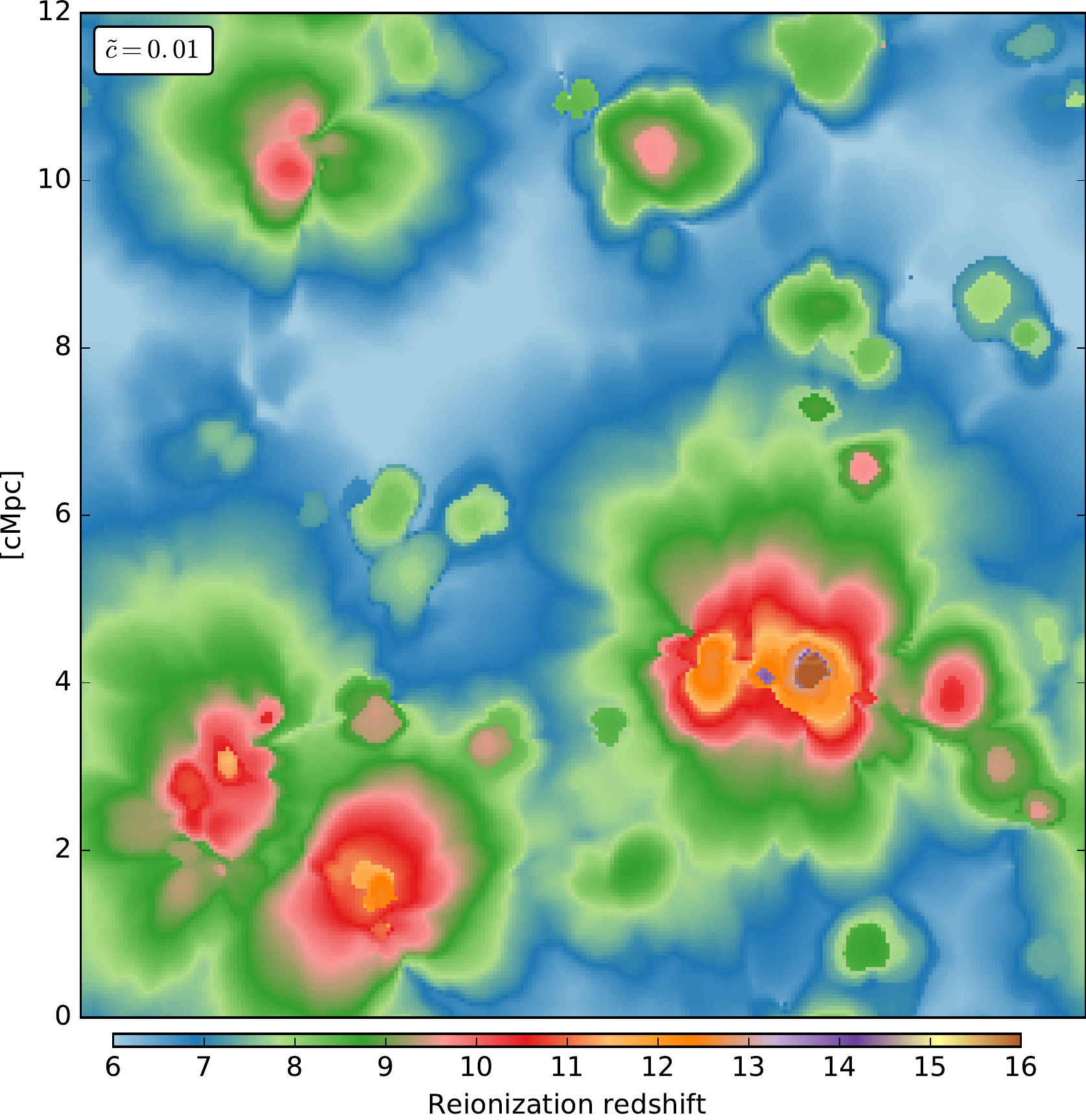}
    \caption{Reionization redshift maps for reduced speeds of light $\tilde c=1,0.1,0.01$.
    \label{fig:xmap}
    }
\end{figure}

\begin{figure}
    \center
\includegraphics[width=0.45\textwidth]{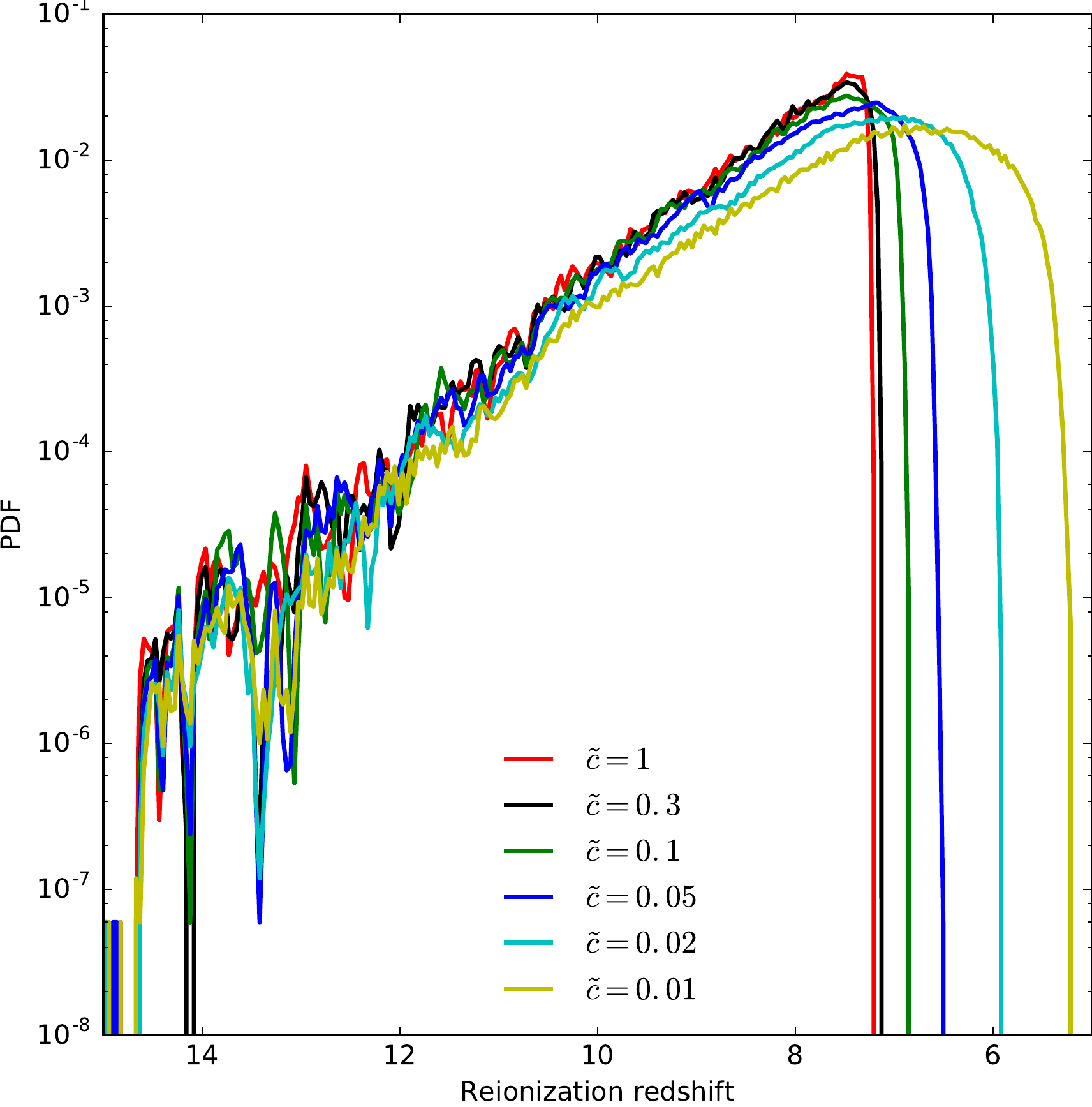}
\caption{Probability density functions of reionization redshifts measured in reionization redshifts maps (see Fig. \ref{fig:xmap}) for different reduced speeds of light.}
 \label{fig:PDFxmap}
\end{figure}

Ionization redshift maps offer an interesting tool for  analyzing the whole reionization process using a single data field. Redshift maps are obtained by keeping in memory, for each cell, the time when its ionization fraction crosses a given threshold. The maps were computed on the fly at each simulation time step to get the highest temporal resolution possible; this is in contrast to similar studies where such maps are computed in post-processing from snapshots with a sparser time sampling (see, e.g., \citet{2013ApJ...777...51O}).

In this study, the cell reionization time is considered to be the first time when the volume weighted ionization fraction goes above 50\%. Figure \ref{fig:xmap} shows three maps of reionization redshifts for three $\tilde c$ values, and the reionization redshift probability density function (PDF) can be found in Fig. \ref{fig:PDFxmap} for all runs.
{Maps are made from slices with a thickness corresponding to one coarse cell ($46$ ckpc)}, taken at the same coordinate in the z-axis for the three simulations.
The slice is chosen to contain the first cell to cross the ionization threshold in the $\tilde{c}=0.1$ simulation.

Qualitatively, the reionization maps for $\tilde c = 0.01, 0.1, 1$ present common features.
The sources are located at the same places, consistent with the fact that the ionizing feedback does not significantly change global star formation processes in our simulations (Fig. \ref{fig:SFRX}).
Radiation escapes high density regions in comparable butterfly shapes {when} it propagates freely in directions that offer less resistance perpendicular to filamentary regions.

The $\tilde{c}=0.1$ run is quite similar to the  $\tilde{c}=1$.
Dark green isocontours ($z\approx6$) are located at the same positions and have the same shape and extension. Greater differences can be noted at smaller redshifts when radiation reaches the underdense regions (in blue); these voids reionize at later times for $\tilde{c}=0.1$ compared to $\tilde{c}=1$, as expected from lower velocities. When the same map for $\tilde{c}=0.01$ is  considered the differences are striking: the underdense regions, distant from the main sites of photon production, are reionized at much later times compared to the two previous experiments with higher velocities. 

The quantitative difference is particularly clear in the redshift {PDFs} (see Fig. \ref{fig:PDFxmap}): at high redshifts all PDFs are similar, but the differences become more notable at later times (starting from $z\sim 11$), and the cumulative effect leads to a significant delay in the global reionization for low $\tilde c$. This is in agreement with the measurements made on the volume weighted ionization history (Fig. \ref{fig:SFRX}): runs with high speed of light reionize {earlier because radiation is able to ionize voids more rapidly. }


%

%
%
%
\subsection{Ionization front speeds}
\label{s:meth}
\begin{figure}
    \center
        \includegraphics[width=0.4\textwidth]{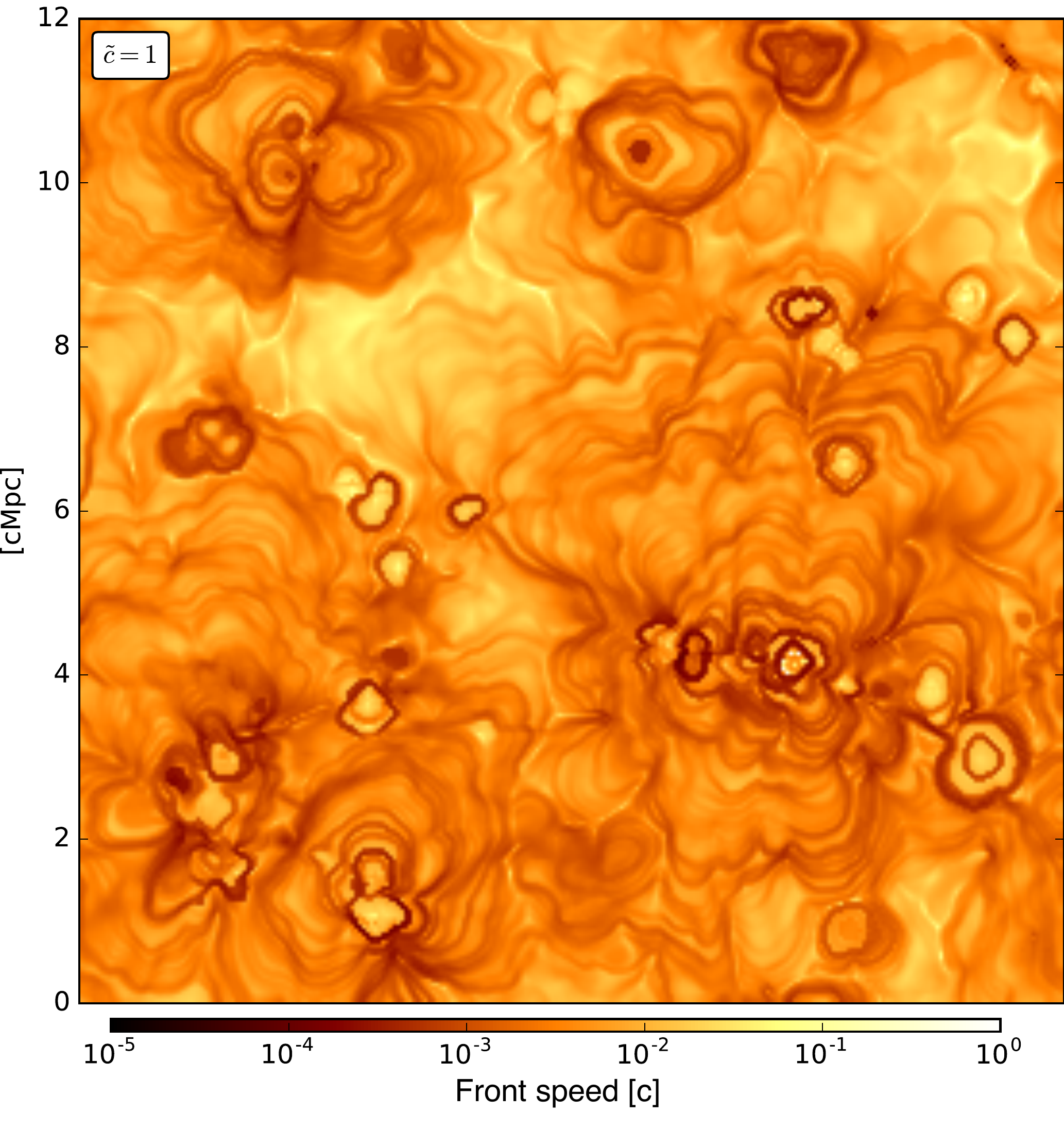}
        \includegraphics[width=0.4\textwidth]{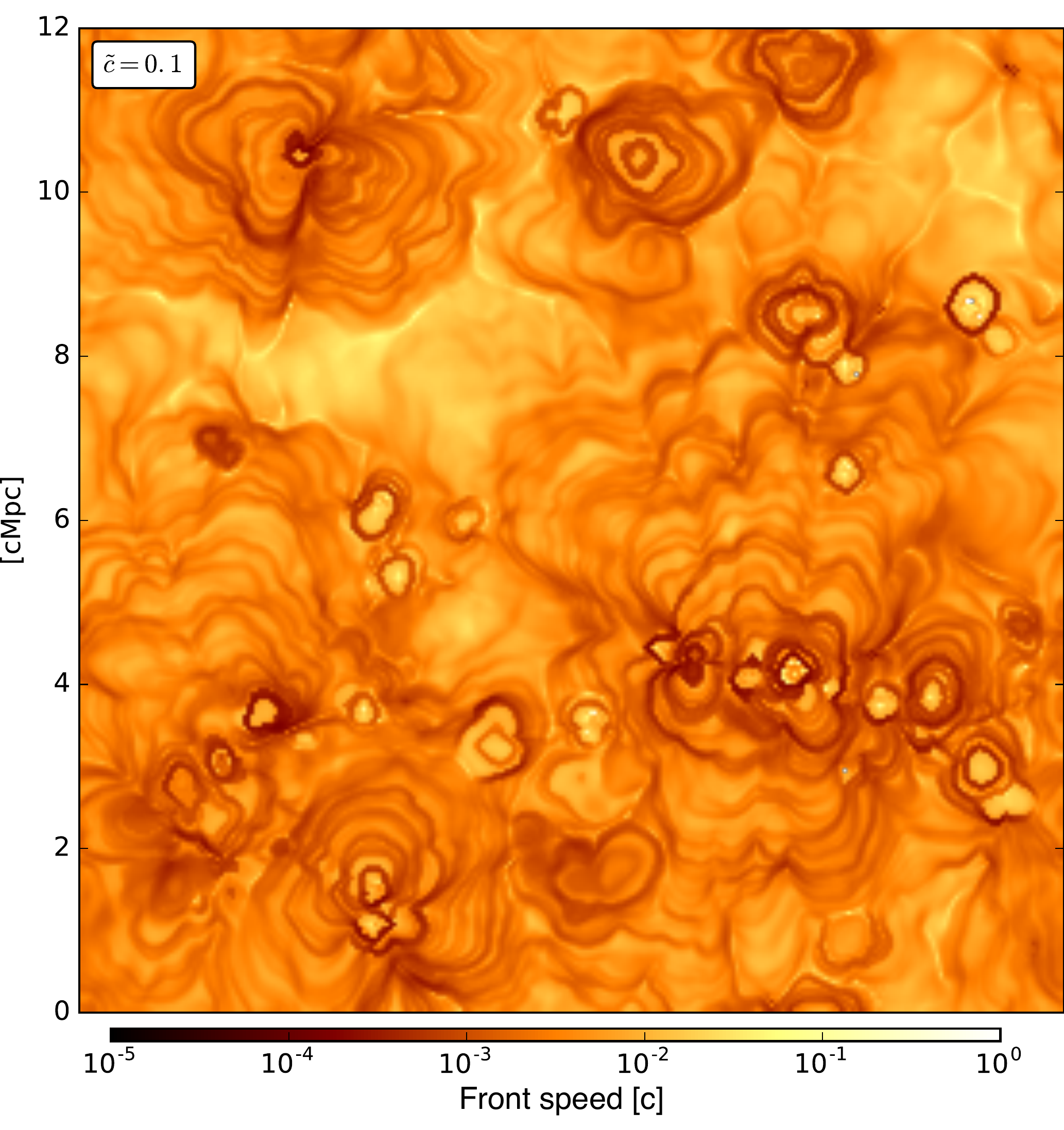}
        \includegraphics[width=0.4\textwidth]{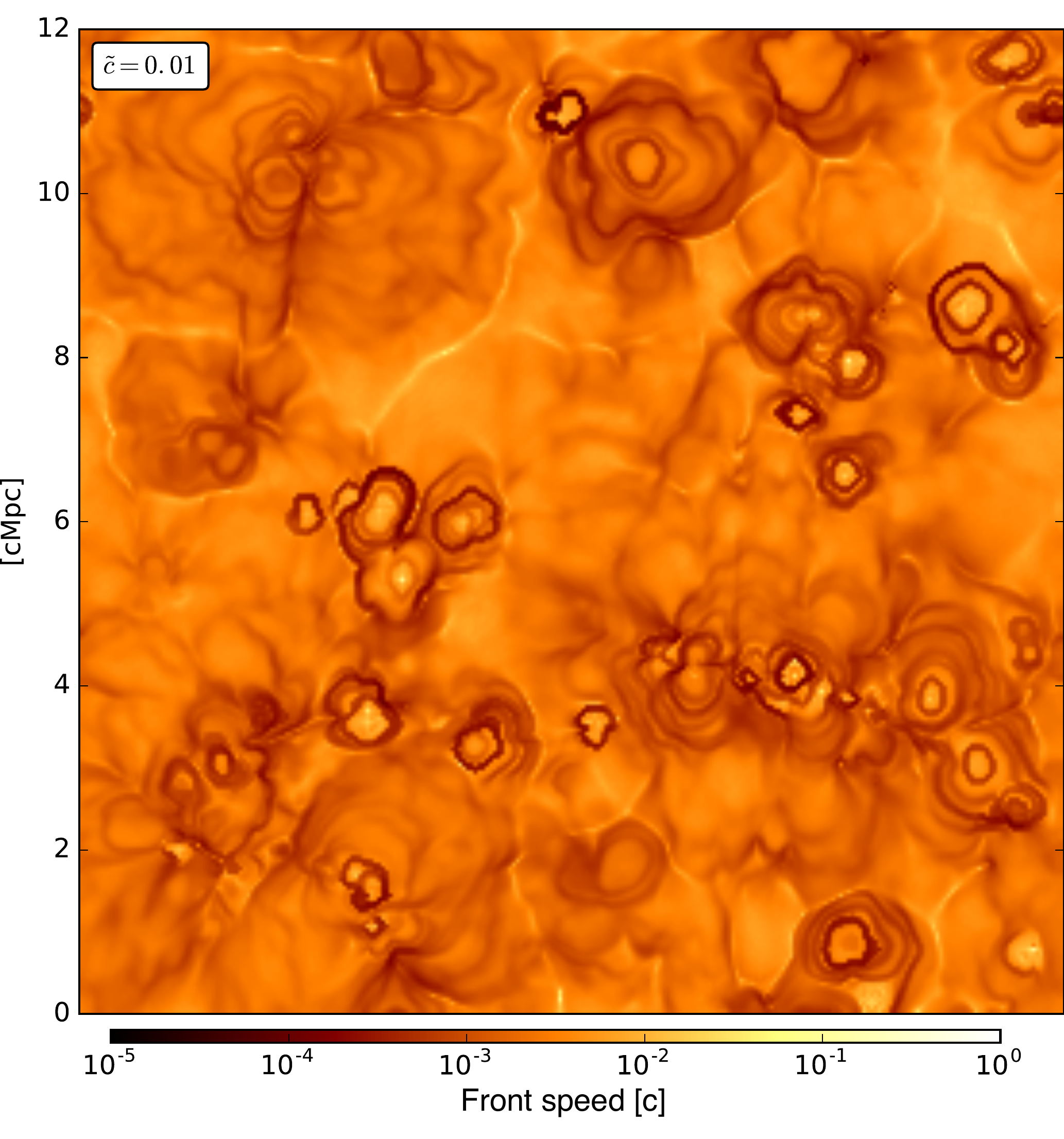}

    \caption{Maps of I-front speed estimates for $\tilde c=1, 0.1, 0.01$. I-front speeds are expressed in units of $c$.
    \label{fig_vmap}
    }
\end{figure}

\begin{figure}
        \includegraphics[width=0.45\textwidth]{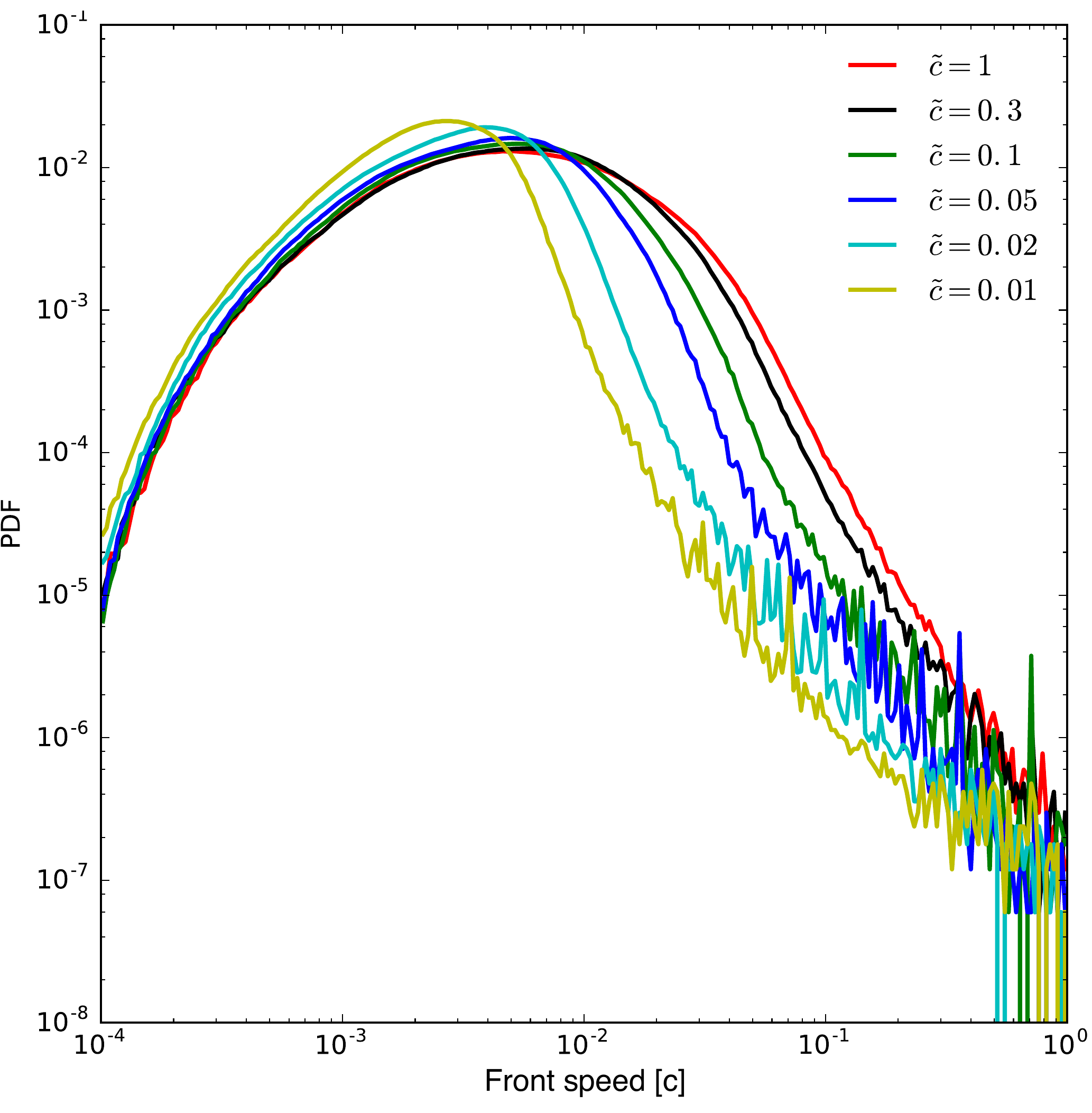}
        \caption{Probability density function of I-front speeds measured in I-front speed maps (see Fig. \ref{fig_vmap}), for different reduced speeds of light.}
        \label{fig:PDFvmap}
\end{figure}

The reionization maps represent a time at each point of the simulation in each cell.
Using the spatial gradient of this map, we can estimate the time required to reionize each cell per unit length. We use a discrete gradient
\begin{equation}
\vec{\nabla} t_{reio}^{i,j,k} = \frac{ \left(t^{i+1,j,k}  - t^{i-1,j,k},t^{i,j+1,k}  - t^{i,j-1,k},t^{i,j,k+1}  - t^{i,j,k-1}\right)}{2\Delta x},
\label{e:grad}
\end{equation}
where $i,j,k$ are cartesian cell indexes and $\Delta x$ is the 1D spatial extent of a reionization map cell. This gradient represents the time needed to ionize a given distance (e.g., in [yr.pc$^{-1}$]), the inverse is therefore equivalent to a velocity (e.g., in [pc.yr$^{-1}$]) :
\begin{equation}
V_{reio}  = \frac{1}{\left |  \vec{\nabla} t_{reio} \right| }.
\label{e:vel}
\end{equation}
Here $V_{reio}$ can be interpreted as an estimator of the speed of the I-front passing through a cell.


%
%

Figure \ref{fig_vmap} shows maps of I-front speeds obtained with this method for three runs with different reduced speeds of light. In the maps presented in Fig. \ref{fig_vmap}, the concentric ``rings'' with high and low speeds fronts are induced by the consecutive events of star formation.
I-fronts can only expand if there are internal radiative sources, but star formation is not continuous and happens in a bursty manner. Fronts slow down during the period between two consecutive star formation events.  {We investigate this effect further in the Appendix.}

It should be noted that the velocity obtained using Eqs. \ref{e:grad} and \ref{e:vel} is only an approximate estimator; it produces an adequate measure in the case of a perfect plane parallel front, although it can be less accurate in the more realistic case of multiple, multi-directional fronts. In particular, limitations appear when neighboring cells have very similar reionization times, more similar than authorized by a single propagating front;  the time gradient in Eq. \ref{e:grad} becomes artificially small and the associated reionization speeds become artificially high and even infinite in the case of neighboring cells with identical reionization times.
Example of such situations are as follows: 
\begin{itemize}
\item In overdense regions where adjacent cells have simultaneous star forming events.
\item In underdense regions where two (or more) I-fronts start to overlap.
\end{itemize}  
As a consequence, measured front speeds can be up to several times greater than the speed of light in the simulation.  In the PDFs of I-front speeds shown in Fig. \ref{fig:PDFvmap}, we observe that limited speeds of light do not prevent the presence of I-front speeds greater than their values. For instance, there are still front speeds of $10^{-1}$ when $\tilde{c}=10^{-2}$, but these values represent a small integrated fraction of the PDFs ($\sim 8\cdot10^{-5}$ at most in all our experiments), thanks in particular to the important time resolution of the reionization maps. Therefore, the probability of  finding velocities greater than $\tilde c$ is low and {the impact of the reduced speed of light on front propagations can be measured using this estimator}. Furthermore, our focus is  on the relative differences of different simulations obtained via  identical methodology rather than measures of absolute values for the velocities.



\subsection{I-front speed as a function of redshift}

\begin{figure}[h]
    \center
        \includegraphics[width=0.37\textwidth]{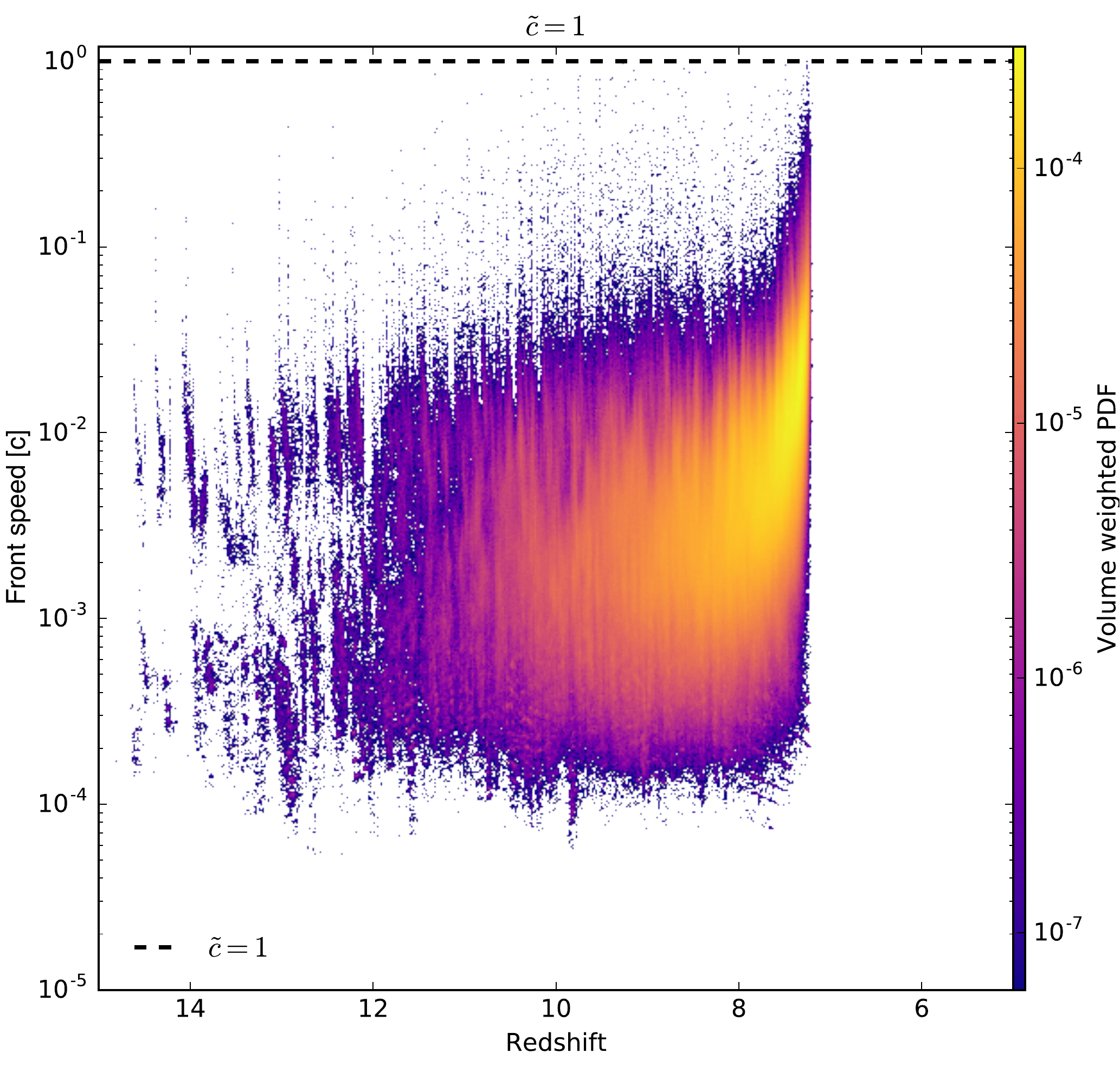}
        \includegraphics[width=0.37\textwidth]{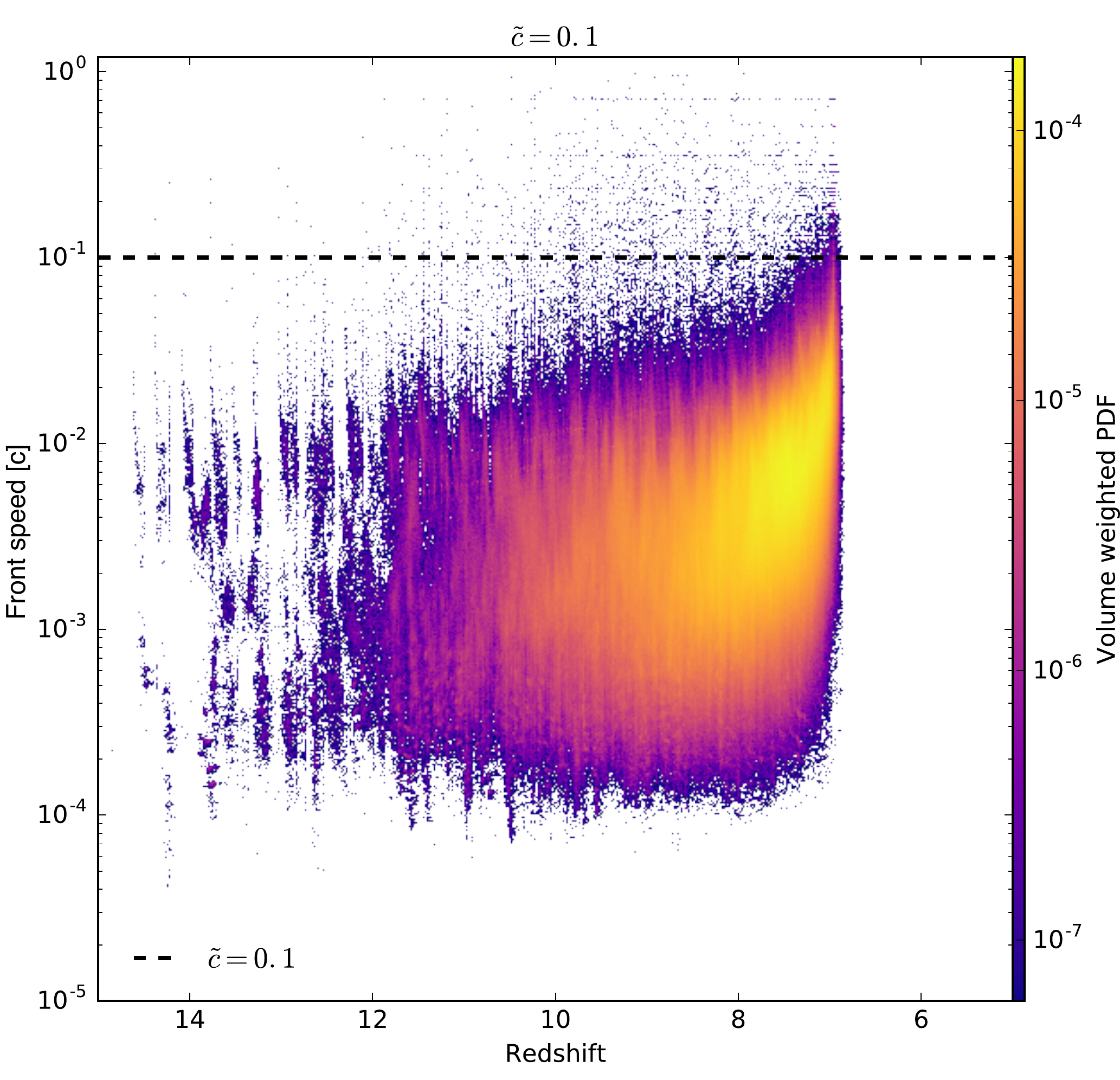}
        \includegraphics[width=0.37\textwidth]{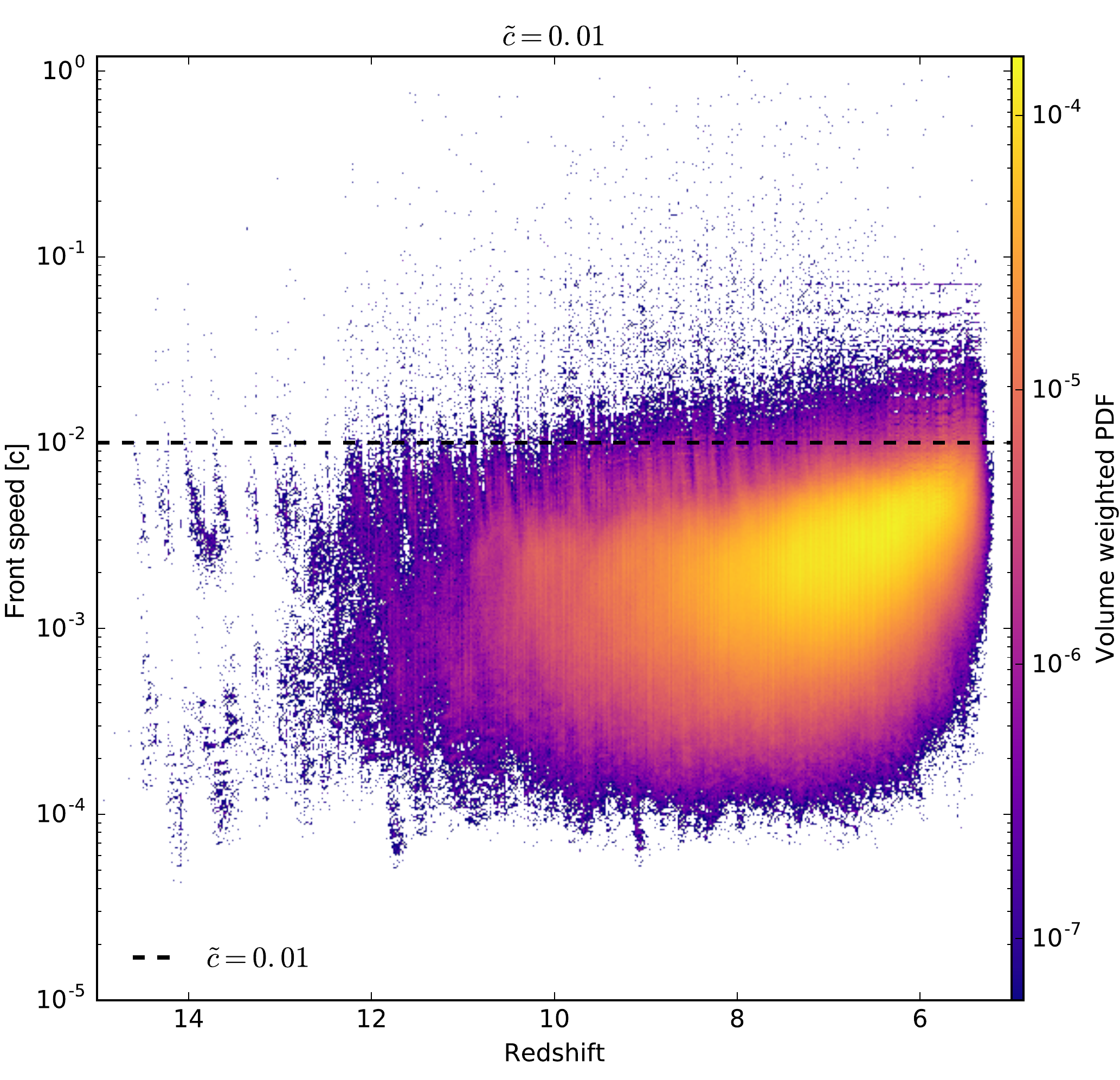}
    \caption{Two-dimensional histograms of  front speeds (expressed in $c$ units) as a function of redshift for $\tilde c= 1,0.1,0.01$.  The horizontal dashed line marks $\tilde{c}$, the reduced speed of light used in the simulation.
    \label{fig_vz}
    }
\end{figure}

\begin{figure}[h]
\includegraphics[width=0.45\textwidth]{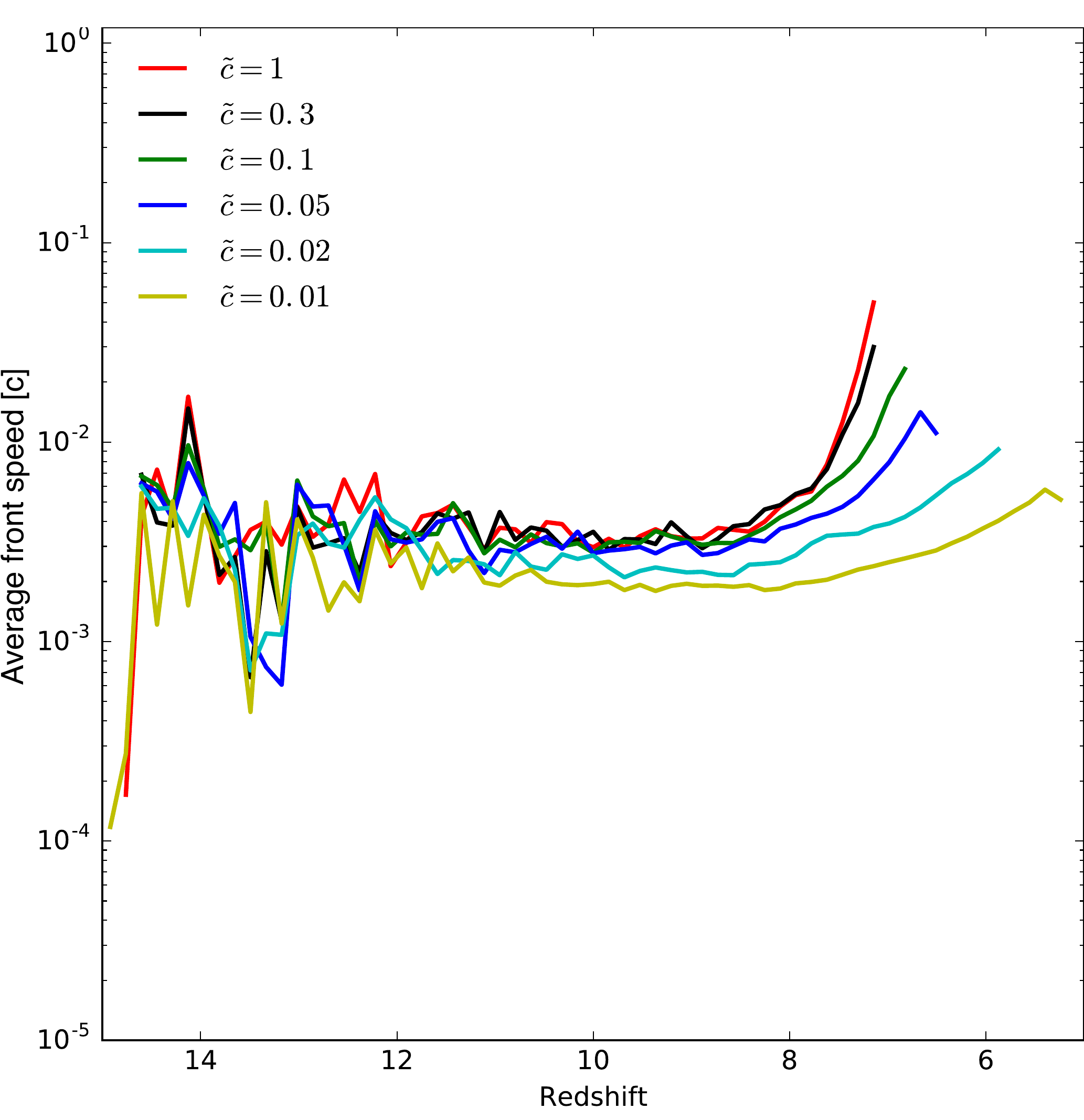}
\caption{Average front speed as a function of redshift for different reduced speeds of light.}
\label{fig:PDFvz}
\end{figure}

The results from the two previous sections can be combined: at each map position we have access to a reionization time/redshift and a front speed. Therefore, front speeds can be binned according to the associated reionization redshift, providing the distributions of front speeds as a function of redshift.
Figure \ref{fig_vz} presents this distribution and Fig. \ref{fig:PDFvz} the evolution of the average front speeds for different $\tilde c$. 

We first focus on the 2D histograms that show the redshift evolution of the distribution of  front speeds  (see Fig. \ref{fig_vz}). In the $\tilde{c}=1$ run, we observe that for z>8, the distribution of speeds remains between $\approx 10^{-1}c$ and $\approx 10^{-4}c$. This phase is then followed by a quick acceleration ``plume'' with velocities close to $c$.
We therefore observe a two-stage process: first, light escapes overdense  regions at an average quasi-constant moderate speed and second, the percolation of ionized bubbles allows radiation to reach underdense regions. In these regions during these later stages, the front speeds are not significantly limited by the interaction with matter. The light is free to fill the voids and front speeds are able to reach high values. 

In the $\tilde{c}=0.1$ run, the first phase is not impacted by the RSLA, which is  expected since the front speeds are already lower than $\approx 10^{-1}c$ for the $\tilde{c}=1$ run.
The main differences appear in the second phase: the $z<8$ peak is reduced by the RSLA,
whereas the $\tilde{c}=1$ experiment presents fronts that are able to reach speeds greater than 0.1c. By setting $\tilde{c}=0.1$, the velocities of these fronts are limited to $\approx 0.1c$ in an artificial manner: the end of the reionization is slightly delayed. At low speed of light values, in the $\tilde{c}=0.01$ run, the differences are even more drastic. Not only are front speeds  severely limited during the late accelerating phase, but a difference can also be seen in the first phase ($z>8$) as extreme front speed values are artificially decreased by an extreme RSLA. It leads to an important cumulative delay in the reionization redshift.

Considering now the average front speeds  as a function of redshift (see Fig. \ref{fig:PDFvz}), we clearly distinguish the two stages: an almost constant value at early times, followed by an acceleration for redshift $z<8$.
In the $z>8$ quasi-constant phase, the average front speed goes from $4.1 \cdot 10^{-3}$ for the  $\tilde{c}=1$ run to $2.2 \cdot 10^{-3}$ for the  $\tilde{c}=0.01$ run.
For reduced speed of light values lower than $\tilde c \sim 0.05$, the impact seems strong enough to shift down the average front speed value compared to other experiments for $z<10$. 
The late-stage acceleration for redshifts $z<8$ is significantly impacted by the RSLA, and as expected a high speed of light value leads to a more sudden ``flash'' of the volume at the end of the reionization than with lower values. This final acceleration corresponds to the light reaching underdense regions, and in all cases the average front speed is limited by a reduced speed of light, even though $\tilde{c}=0.3$ is quite consistent with $\tilde{c}=1$. To sum up, reduced speed of lights $\tilde{c}>0.05$ can be considered as satisfying {to reproduce}  the first ``slow'' stage at quasi-constant speed for front propagations, but will likely induce a delay for the overlap if $\tilde{c}<0.3$.

These findings also provide additional insights into the influence (or lack of influence) of RSLA as a function of local matter distribution by relating these two stages to two different regimes of matter density. The first phase is linked to the initial stages of the production of radiation by the first sources in dense regions. In this regime the RSLA can be a good approximation down to $\tilde{c}=0.05$. By extension, we can assume that speeds of light as low as this value can be used to study overdense objects at any time, for example to study the escape fraction of ionizing radiation of galaxies. The second phase with high  I-front speeds is related to the overlap of HII regions as radiation propagates in the more diffuse intergalactic medium (IGM) or in voids: in this regime of densities, when radiation propagates freely, I-front speeds can be as high as $c$ and a reduced speed of light is always a limiting factor. For instance, the remote influence of one galaxy on another distant one, via the propagation of radiation through diffuse matter, is likely to be improperly timed at any time. 

Numerical parameters such as the box size or the spatial resolution are likely to impact our quantitative predictions, due to this local density influence. Increasing the resolution should lead to denser regions and potentially slower I-fronts in the first phase. Increasing the box size provides greater underdense regions or voids and could induce faster I-fronts in the second accelerating phase. Overall, further investigations are needed to assess the density or environmental dependence of I-front speeds and how the RSLA impacts the prediction of simulations: we initiate these investigations in the next section.

\section{Probing different regimes of volumes and densities}
\label{s:sim64}

{In order to probe different regimes of volumes or densities, we produced an additional set of reionization simulations with a larger volume $(64 \mathrm{Mpc/h})^3$ and a lower mass resolution ($512^3$ dark matter particles with a $2\times 10^8 M_\odot$ mass and $512^3$  cells in the initial grid + 5 refinement levels). Reionization maps and front velocities were computed in an identical fashion, thus providing front characteristics in regimes of greater volumes and lower resolution. Light speeds $\tilde c=1$ and $\tilde c=0.1$ were used, bracketing the regimes where convergence was achieved in the previous smaller simulations.  The stellar emission was set to reionize the full volume by z=6 for the $\tilde c=0.1$ model;  all the other parameters were kept identical to the main set of models discussed previously. With this setup, the $\tilde c=1$ simulation is reionized at z=7;  both simulations $(64 \mathrm{Mpc/h})^3$ reionize later than their small box equivalents using the same $\tilde c$ values. 
}
\begin{figure}[h]
\includegraphics[width=0.45\textwidth]{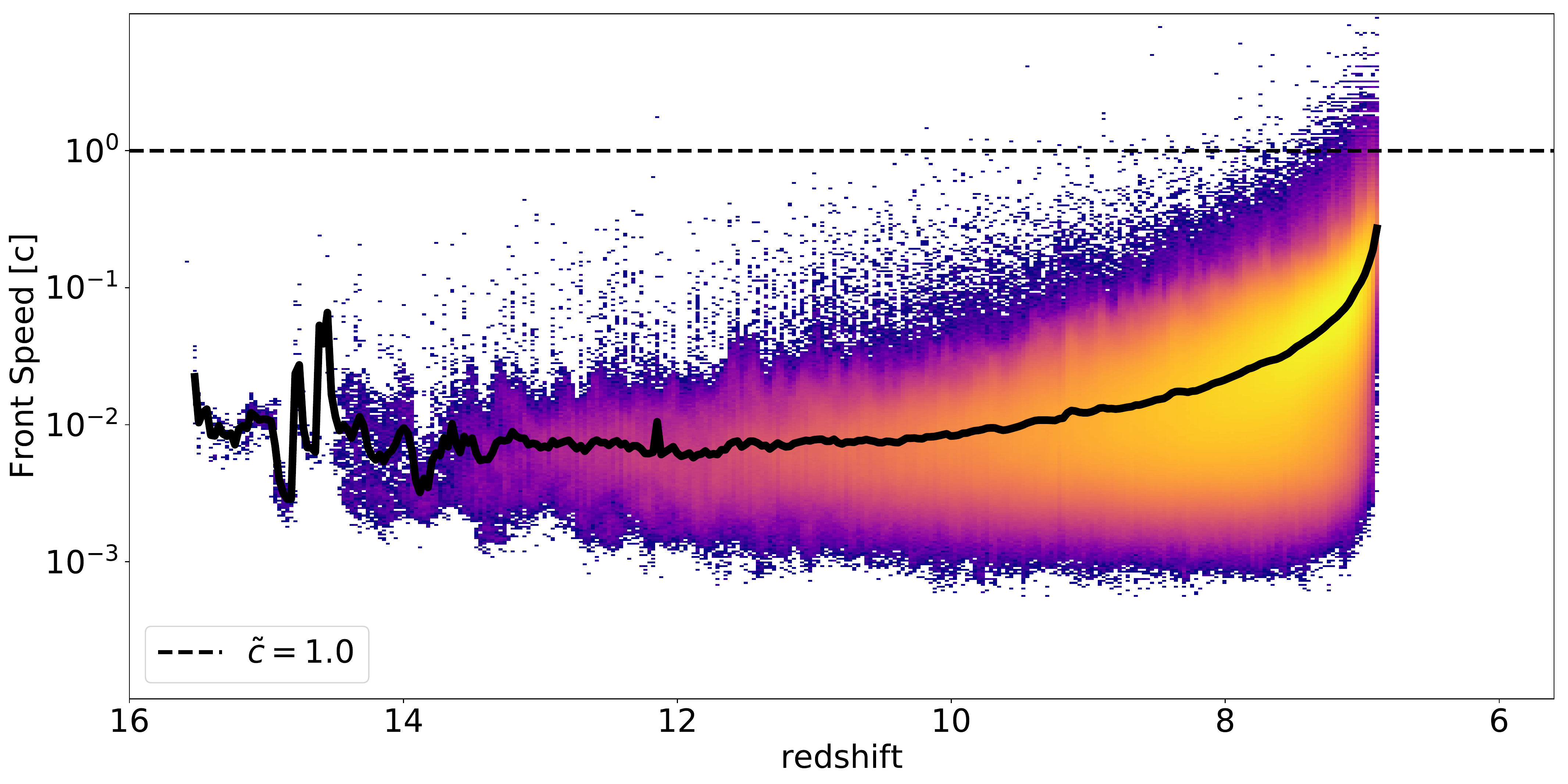}
\includegraphics[width=0.45\textwidth]{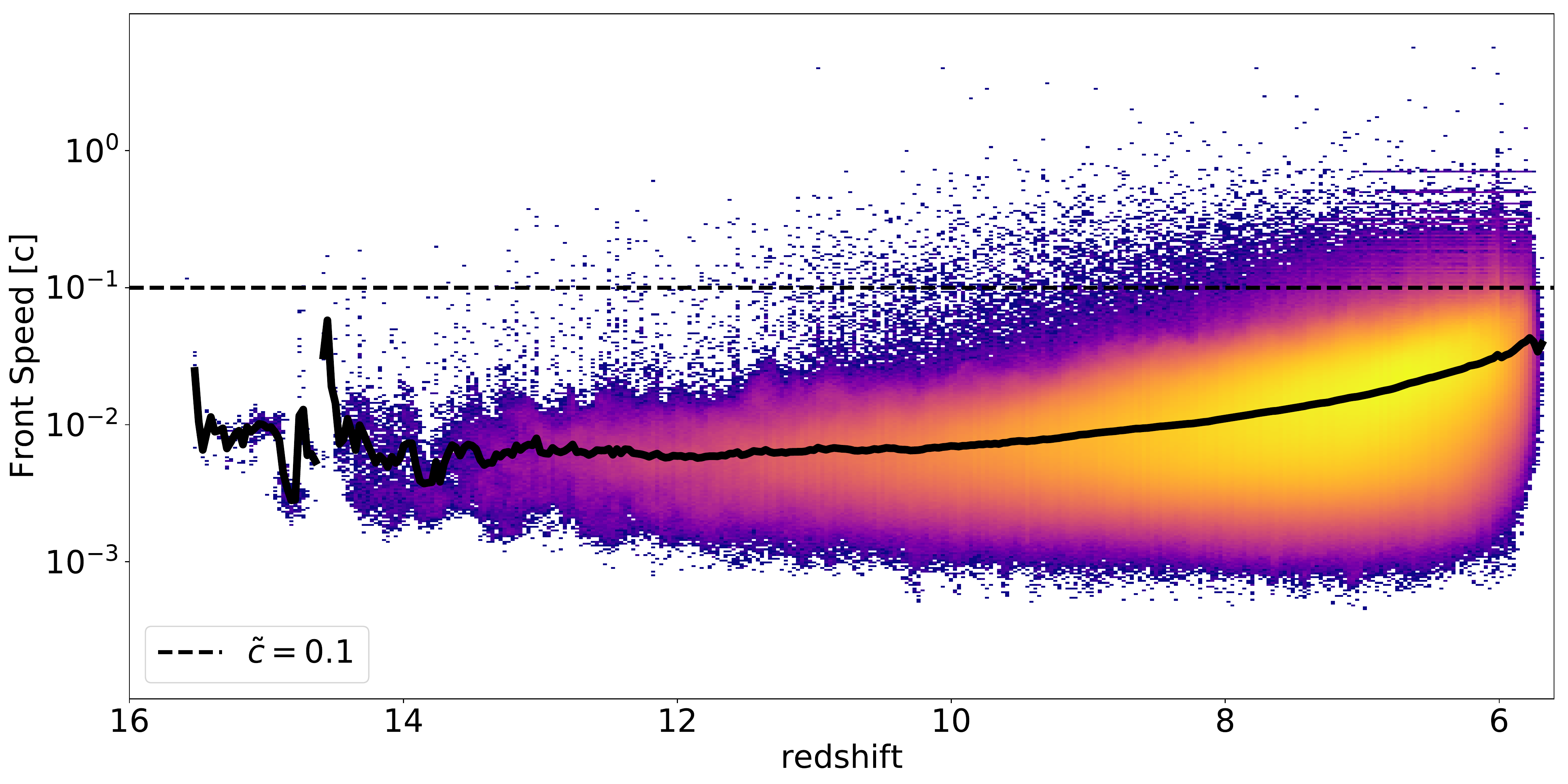}
\caption{Two-dimensional histograms of  front speeds (expressed in $c$ units) as a function of redshift for $\tilde c= 1\ \mathrm{and}\ 0.1$ (upper and lower panel, respectively) in large 64 Mpc/h volumes. In both cases the bold solid lines stand for redshift evolution of the average velocity.}
\label{fig:histvel64}
\end{figure}

{The redshift evolutions of the front speed distributions are shown in Fig. \ref{fig:histvel64}. Overall, this different regime of volume and resolution leads to a global behavior similar to that encountered previously: velocities evolve according to a two-stage process with an initial slow-front regime followed by an increase as the overlap of HII regions take place. In both cases the average front velocities result in values close to the speed of light, and in both cases values higher than $\tilde c$ can be measured, due to the same limitations of the methodology discussed previously (see Sect. \ref{s:meth}). Also, the $\tilde c=1$ model experienced an earlier reionization compared to the $\tilde c=0.1$ model, as already observed for the smaller simulations (see Fig. \ref{fig:SFRX}).
}

{One notable difference compared to the small volume models can be noted in the initial slow front stage: velocities are higher in the current, less resolved simulations.  Notably, local velocities as low as $10^{-4} c$ cannot be measured in the two simulations, whereas such values were found in the fiducial set of smaller simulations with higher mass resolution. Likewise, the typical average front velocity is $\sim 10^{-2}$, higher than previously found.  This regime is driven by the densest regions around the first sources and the lack of high density contrast in these large volume simulations allows the fronts to experience less ``resistance'' as they escape into the more diffuse  IGM.
}

{At later stages close to the overlap, front velocities tend to be higher for both values of the speed of light compared to the simulations made in smaller volume. In fact, spurious values greater than $\tilde c$ are more frequent in both cases, even though they still represent a small fraction of the total number of cells ($< 0.001\%$ in both cases). High values of front velocities, close to $\tilde c$, are found at earlier times compared to the smaller boxes, even though reionization takes place later;  the late stage evolution is shallower and less sudden in the 64 Mpc/h simulations than models produced in 8 Mpc/h boxes. Large underdense regions allow fast propagation at larger redshifts through a more diffuse IGM.
}

{Thanks to the 64 Mpc/h volume probed here, we can also assess the influence of external sources on the previous small-box experiments. For both $\tilde c$ values, we split the $(64 \mathrm{Mpc/h})^3$ volumes into $8\times 8 \times 8=512$ smaller $(8 \mathrm{Mpc/h})^3$ cubes. Each of these smaller subvolumes would experience a reionization driven by the properties of its own sources, but also influenced by external fronts created by external sources.}
\begin{figure}
\includegraphics[width=0.45\textwidth]{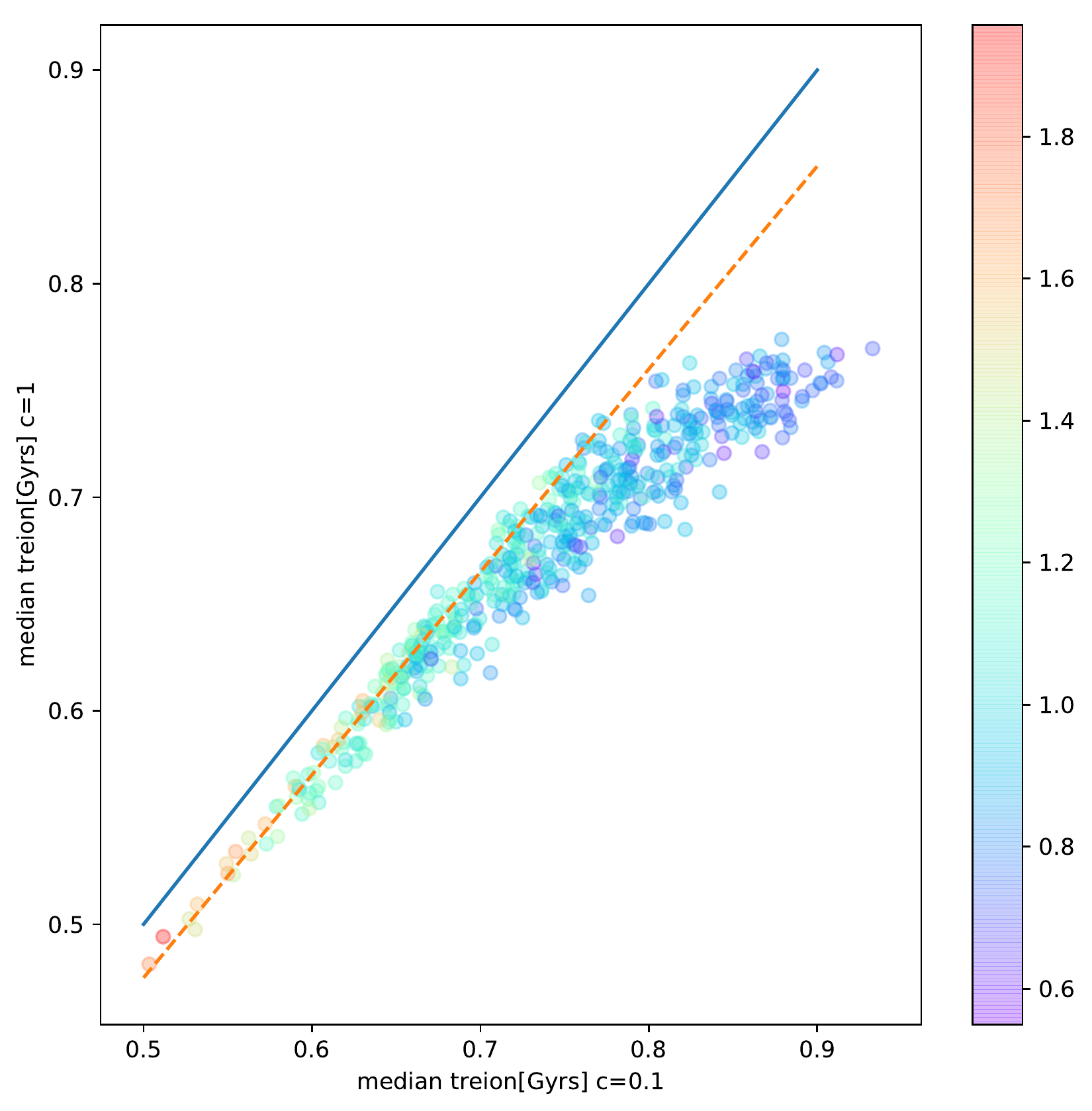}
\includegraphics[width=0.45\textwidth]{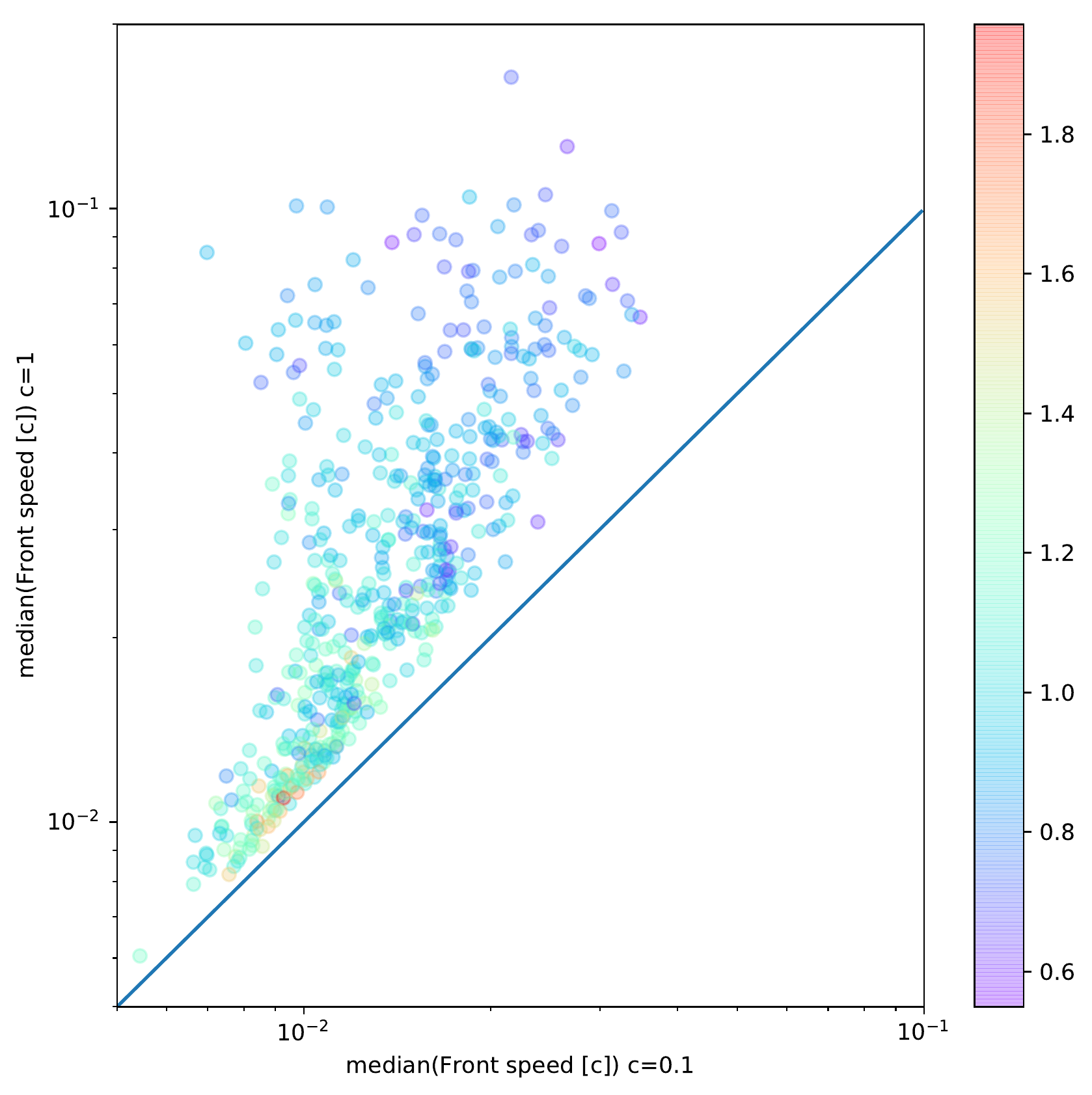}
\caption{Comparison of median reionization times (top) and median front velocities (bottom) obtained in 64 Mpc/h simulations with $\tilde c=1$ and $\tilde c=0.1$. Each dot represents a (8 Mpc/h) subvolume, color-coded as a function of its baryonic average overdensity at z=6. In both panels solid lines stand for the 1:1 correspondence. In the top panel, the dashed line stands for the same correspondence shifted down ($\times 0.95$) to fit the trend at early times.  }
\label{fig:comp64}
\end{figure}
{Since both $\tilde c=1$ and $\tilde c=0.1$ simulations share the same set of initial conditions, such subvolumes can be compared from one simulation to another. The comparisons are shown in Fig. \ref{fig:comp64}. First, we compare the median reionization time of each subvolume for both values of the speed of light. These times present a strong correlation: the volume that reionizes the earliest in one simulation also reionizes at early times in the other. The progression of the reionization is driven by the buildup of structures and sources, a process only weakly modified by the speed of light. In addition, it should be  noted that the densest regions (the density of each subvolume being color-coded) are reionized first, as expected if sources tend to  appear first in these regions, whereas the most underdense are reionized last. Despite the tight correlation, an offset from the 1:1 correspondence can be seen between the subvolume reionization times of the $\tilde c=1$ and $\tilde c=0.1$ simulations. This offset is not systematic, as can be seen when comparing the two simulations on a cell-by-cell basis (see Appendix). It  results instead from the presence of regions that experience a delayed reionization in all the (8 Mpc/h) subvolumes. 

In addition, a departure from a linear correlation can be seen for the volumes that reionize last, corresponding to the volumes that are the most underdense. In these cases, the reionization delay of the reduced speed of light simulation is  more significant than that observed in denser subvolumes. In the bottom panel of Fig. \ref{fig:comp64} we show the comparison of median front velocities{, taken from the distribution of velocities at all times,} in the two simulations. Volumes that reionize the earliest (and therefore the densest ones) present median front speeds that differ only slightly when comparing the $\tilde{c}=1$ and $\tilde{c}=0.1$ experiments. On the other hand, volumes that reionize the latest present a larger departure from the 1:1 correspondence; underdense regions promote high velocities through a diffuse IGM, and are therefore more sensitive to the choice of $\tilde c$.

{This large sample of subvolumes gives us some insight into the impact of cosmic variance, finite-volume effects, and external sources. Dense regions are likely to host sources at early times that create slow fronts, and such volumes are self-reionized. Another possibility could be that these regions are swept by slow fronts created in even denser, nearby regions. In this regime of density, we find that a $\tilde c=0.1$ speed of light has only a moderate impact on front velocities and reionization times. On the other hand, underdense regions are reionized at later times and are therefore likely to be externally reionized by sweeping fronts; in a reduced speed of light regime, such fronts would arrive at later times and be slower. In such cases, a high speed of light would be necessary. The transition between these two regimes takes place approximately at the average cosmic density that was enforced by construction in our previous set of 8 Mpc/h simulations; a conservative stance would therefore push for a high $\tilde c$ if external fronts have to be taken into account in this specific case. In any case, and as demonstrated here, effects of cosmic variance, finite volume, or density regimes can be quite important in general.
}

\section{Conclusions}

We described a method used to estimate ionization front speeds from reionization maps produced by cosmological simulations. With this method we find that reducing the speed of light modifies the propagation of fronts in our simulations.

We find that the propagation of radiation in our models of reionization is a two-stage process:  a first phase with quasi-constant, moderate average I-front speeds, and a second stage with  an acceleration phase during which the average I-front speed increases greatly, close to the speed of light.  {Using a set of small $(8 \mathrm{Mpc/h})^3$ experiments we found the following:}
\begin{itemize}
\item During the early constant speed phase, the average value does not depend on $\tilde{c}$ for $\tilde{c}>0.05$. 
\item The second acceleration phase is always impacted by a reduction in the speed of light, but convergence can be achieved for $\tilde c \sim 0.3$ to reasonably reproduce the redshift evolution of average I-front speeds. 
\end{itemize}
{These quantitative prescriptions are likely to depend on the type of simulation (resolution, volume, external sources). In order to get some insight on the robustness of these results, we performed the same analysis on a set of similar $(8 \mathrm{Mpc/h})^3$ volumes extracted from a larger $(64 \mathrm{Mpc/h})^3$ simulation with a lower mass resolution. We find that  
in the slow phase there are  greater front speeds,   but still well below $0.1 c$, and that 
 large variations in median front velocities and reionization times can be found  in underdense regions that experience late reionization when comparing the $\tilde c=1$ and $\tilde c=0.1$ experiments. The influence of fronts created by distant sources is therefore likely to be impacted by the choice of speed of light.

Overall, our finding shows that quantitative prescriptions on the speed of light can be quite sensitive to the experimental setup. In addition the inclusion of bright sources creating large-scale fluctuations in the radiation field, such as quasars (see, e.g., \citet{2015MNRAS.453.2943C}) or X-rays with large mean free paths (see, e.g., \citet{2010A&A...523A...4B}), may change the quantitative results obtained here on front velocities.  } {We also did not consider the impact of an additional homogeneous radiation background, which  could affect propagation in the late-reionized, low density, regions, i.e., the regimes where the greatest discrepancies are being measured for different values of $\tilde c$. Even though this lack of background is consistent with many previous studies made on such scales where the reionization is driven by the population of simulated sources, such a background could  dramatically lower the impact of reduced speed of light by controlling the ionization state of vast regions. Our conclusions should therefore be considered in the strict, but widely accepted context of reionization models driven by the ``simulated sources''.}

{The two-stage evolution of front velocities is likely to be a robust finding, as demonstrated for example by recent similar findings by \citet{2018arXiv180709282D}, and could potentially be used to define domains of validity of reduced speed of light experiments, depending on the sizes, densities, or redshifts probed. However, beyond its impact on the propagation of fronts, nontrivial effects of the reduced speed of light could exist, for example on  the photo-suppression of distant star formation, the relative delays between different kind of feedbacks (see, e.g., \citet{2017MNRAS.470..224T}), or the gas thermochemistry (see, e.g., \citet{2018arXiv180302434O}). Systematic studies have yet to be generalized.}



%
\bibliographystyle{aa} 
\bibliography{biblio.bib} 
%

\begin{acknowledgements}
      This work is supported by the ANR ORAGE grant ANR-14-CE33-0016 of the French Agence Nationale de la Recherche. This work was granted access to the HPC resources of CINES under the allocation 2019-A005041061 made by GENCI. The authors thank the referee for the constructive remarks and Harley Katz for the discussions.
\end{acknowledgements}
\begin{appendix}
\section{Impact of the speed of light on the front speed spatial distribution.}
\begin{figure}
\includegraphics[width=0.5\textwidth]{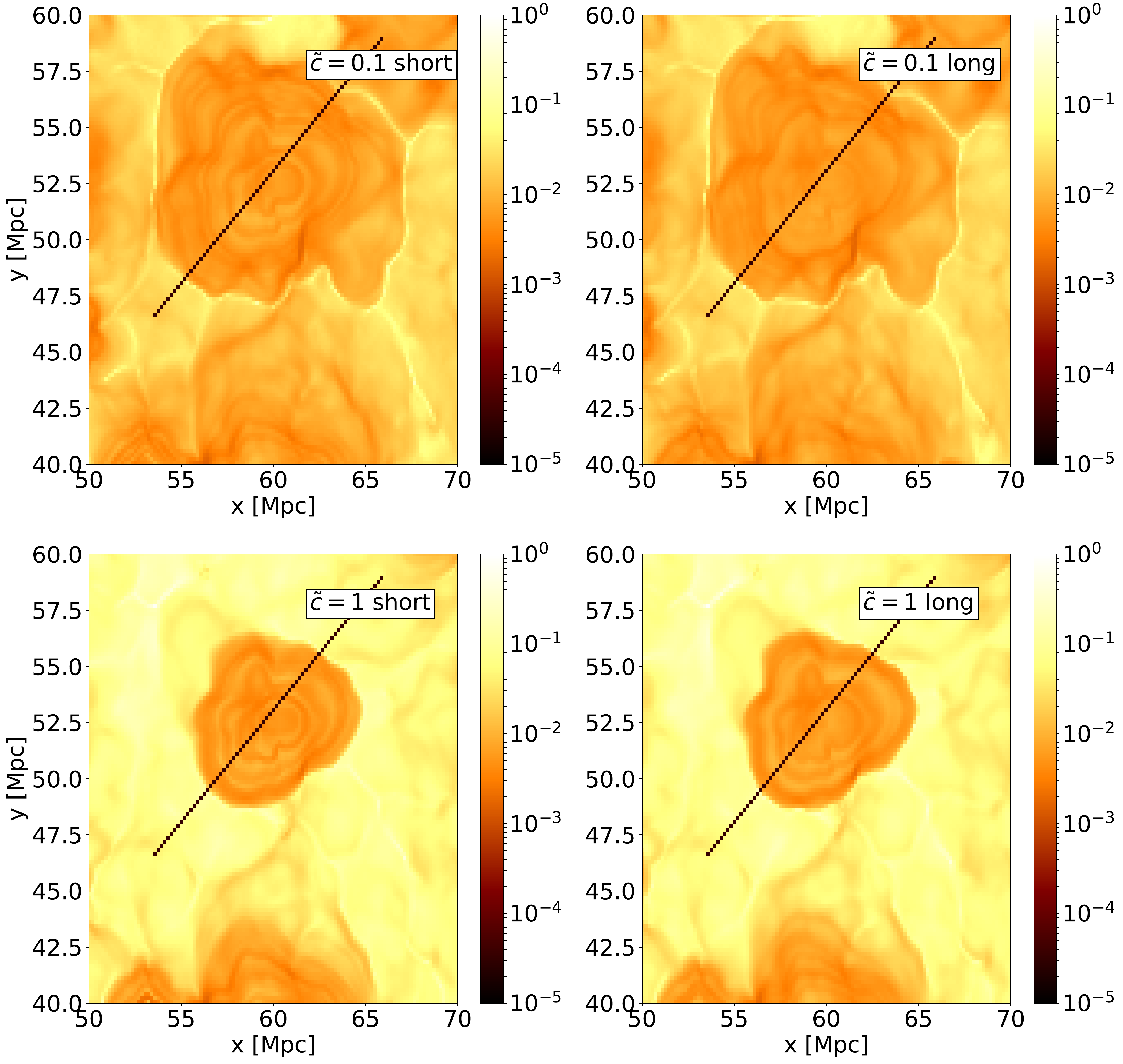}
\caption{Front velocity field in 64 Mpc/h reionization simulations around a source.  Shown are simulations made with $\tilde c=0.1$ (top) and   $\tilde c=1.0$ (bottom). In both cases, two experiments were performed, one with a bright short-lived source, the other with a fainter long-lived source (left and right column, respectively). Solid lines stand for positions where the profiles of Fig. \ref{fig:compstar_profile} are computed}
\label{fig:compstar}
\end{figure}

\begin{figure}
\includegraphics[width=0.5\textwidth]{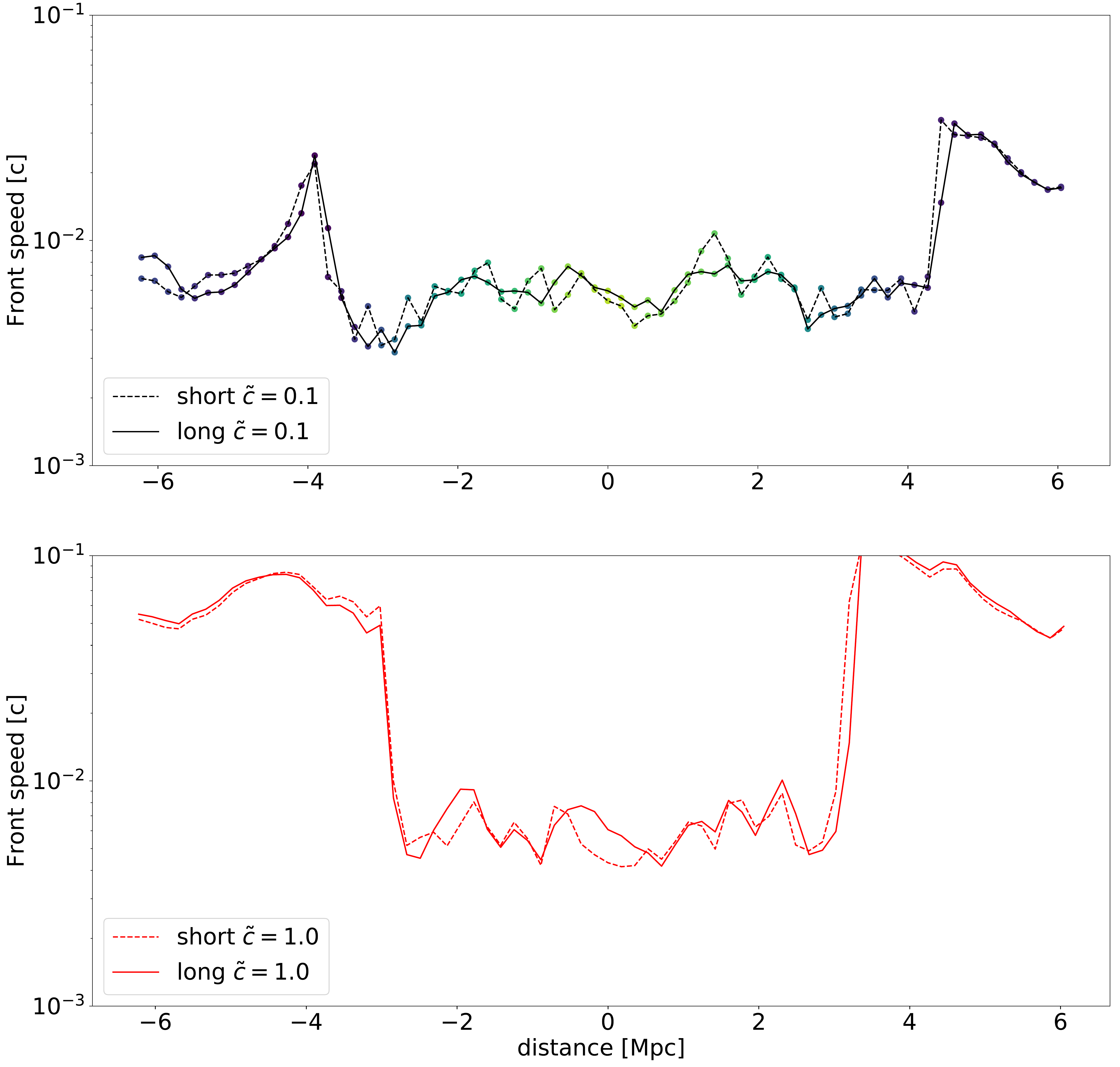}
\caption{Values of the front velocity field in 64 Mpc/h along the solid lines shown in Fig. \ref{fig:compstar}. Shown are profiles taken from $\tilde c=0.1$ simulations (top) and  $\tilde c=1$ simulations (bottom).}
\label{fig:compstar_profile}
\end{figure}

{As discussed in the main text, ripple-like features can be seen in the spatial distribution of front speeds, which we attribute to imprints of the episodic nature of photon production by the sources. In this Appendix, we briefly discuss this effect and how it may be affected by reduced values of the  speed of light.}

{In Fig. \ref{fig:compstar} we present four different experiments made using the same setup as  described in Sect. \ref{s:sim64}, consisting of reionization simulations in a large volume (64 Mpc/h, $512^3$ base resolution), with $\tilde{c}\sim 1$ and $\tilde{c}\sim 0.1$. Each pair of simulations was run using two different recipes of photon production from the stellar particles. The first recipe follows the Starburst 99 model described in section \ref{s:sim64}, leading to a steady production rate of ionizing photons for $\sim 3$ Myrs before a sharp decline. The second recipe spreads the initial steady production rate of photons over 10 Myrs, while keeping the same total amount of produced photons. Therefore, we end up with a set of four simulations consisting of two values for the speed of light and two types of photon production: ``short and bright'' and ``long and faint''. Figure \ref{fig:compstar} presents the front velocity spatial distribution around the same source in the four different setups. Figure \ref{fig:compstar_profile} shows front velocity profiles extracted from the same distributions.}

{In the four configurations the typical ripples can be observed in the velocity field. At large distances from the source, fronts can achieve greater speeds in the $\tilde c=1$ simulations, presented as lighter regions on the maps, as expected. If we first consider the reduced speed of light experiment, $\tilde c=0.1$, the episodic nature of the photon production leaves a different imprint, depending on its parameters. The short and the long source scenarios present the same overall spatial distribution of front velocities, however the ripples are smeared out in the case of a ``long and faint'' production of photon. Conversely, the ``short and bright'' model presents a greater number of ripples with more contrast; the bursty episodes of photon production can be detected in the structure of the velocity field. The difference is particularly striking in the innermost part of the ionized region. Meanwhile the same innermost parts  in the $\tilde c=1$ experiments do not seem to be impacted by the difference in photon production history, and present very similar profiles for both types of sources (see Fig. \ref{fig:compstar_profile}), which  suggests that for this peculiar source the delay between two trains of photons is greater than 10 Myrs, because otherwise the ripples would be smeared out.}

{However, it can be seen that velocities are quite similar in the four experiments at small distances (see Fig. \ref{fig:compstar_profile}), with a measured velocity close to $0.005 c < \tilde c$, and is therefore independent of the choice of $\tilde c$. Ripples in the velocity field are fluctuations around low values rather than strong peaks. This suggests that the visibility of the ripples, which in all cases are seen in the slow progression regime,  is not related to differences in long-distance propagations. }

{Overall, in the current experiments, the differences induced by the source models remain marginal and do not have a strong impact on reionization times distributions. We do not exclude that there might be regimes where the cumulative effects of these local differences in front velocities could lead to strong divergence in the front propagation on large scales, for different models of photon production, and when reduced values of  $\tilde c$ are used, which  would require specific experiments to be demonstrated. However, since most of the observed differences take place in the slow-front regime, we think this lack of convergence is unlikely for reasonable stellar models.}

\section{Cell-by-cell comparison of reionization times and front velocities with different speeds of light}
{In this section we briefly describe the raw data that led to  Fig. \ref{fig:comp64}. Instead of showing the reionization times and front velocities averaged in (8 Mpc/h)$^3$ boxes extracted from large (64 Mpc/h)$^3$ simulations, Fig. \ref{fig:comp_cell} compares the same quantities on a cell-by-cell basis for the $\tilde c =0.1$ and $\tilde c=1.0$ large simulations. In particular, it should be noted that no systematic offset from the 1:1 correspondence is observed;  the tendency of quantities averaged over a (8 Mpc/h)$^3$ to drift away from this correspondence in Fig. \ref{fig:comp64} is simply due to the systematic presence of underdense regions within that experience delayed reionizations when $\tilde c=0.1$ is used.}

\begin{figure}
\includegraphics[width=0.45\textwidth]{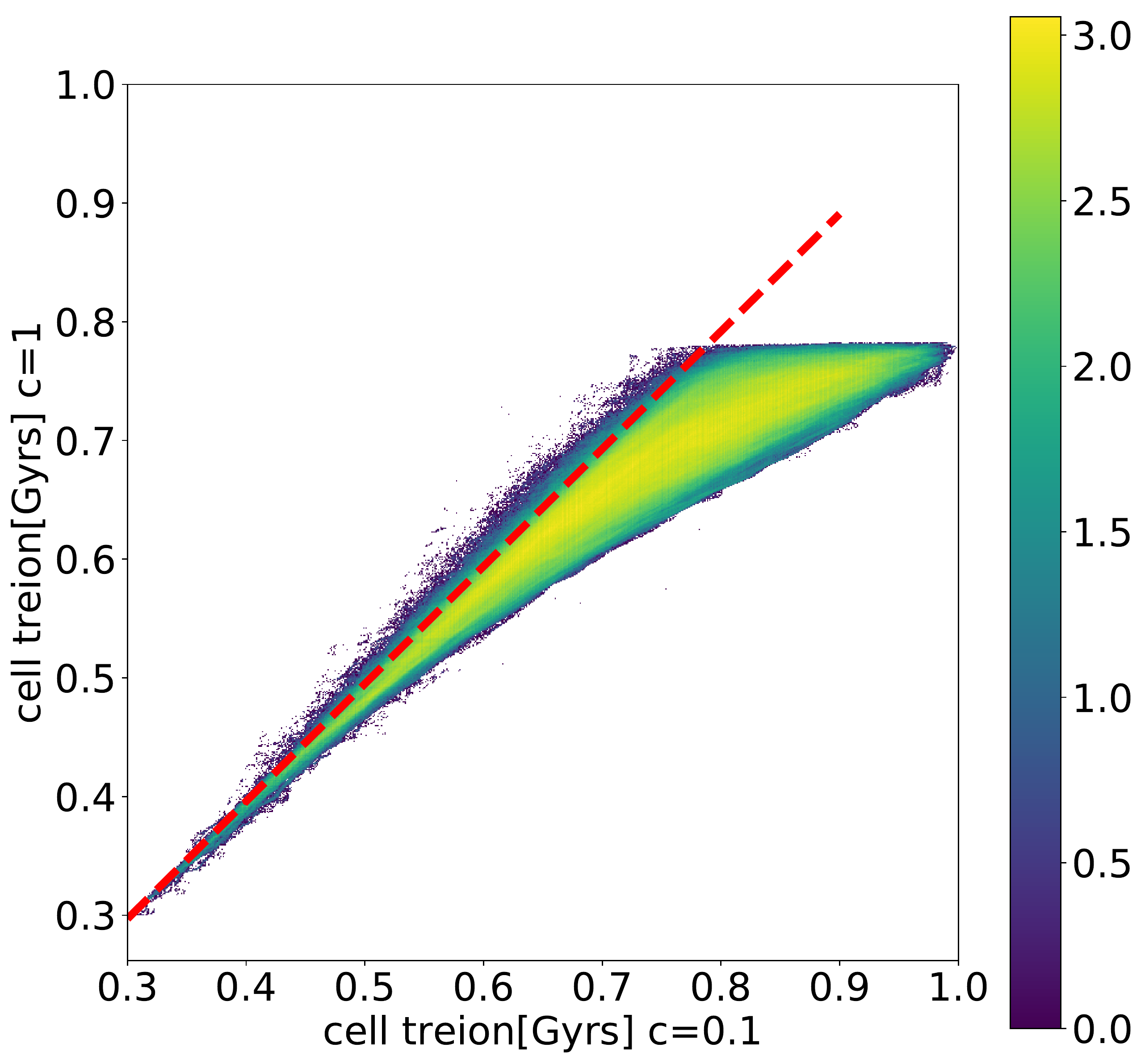}
\includegraphics[width=0.45\textwidth]{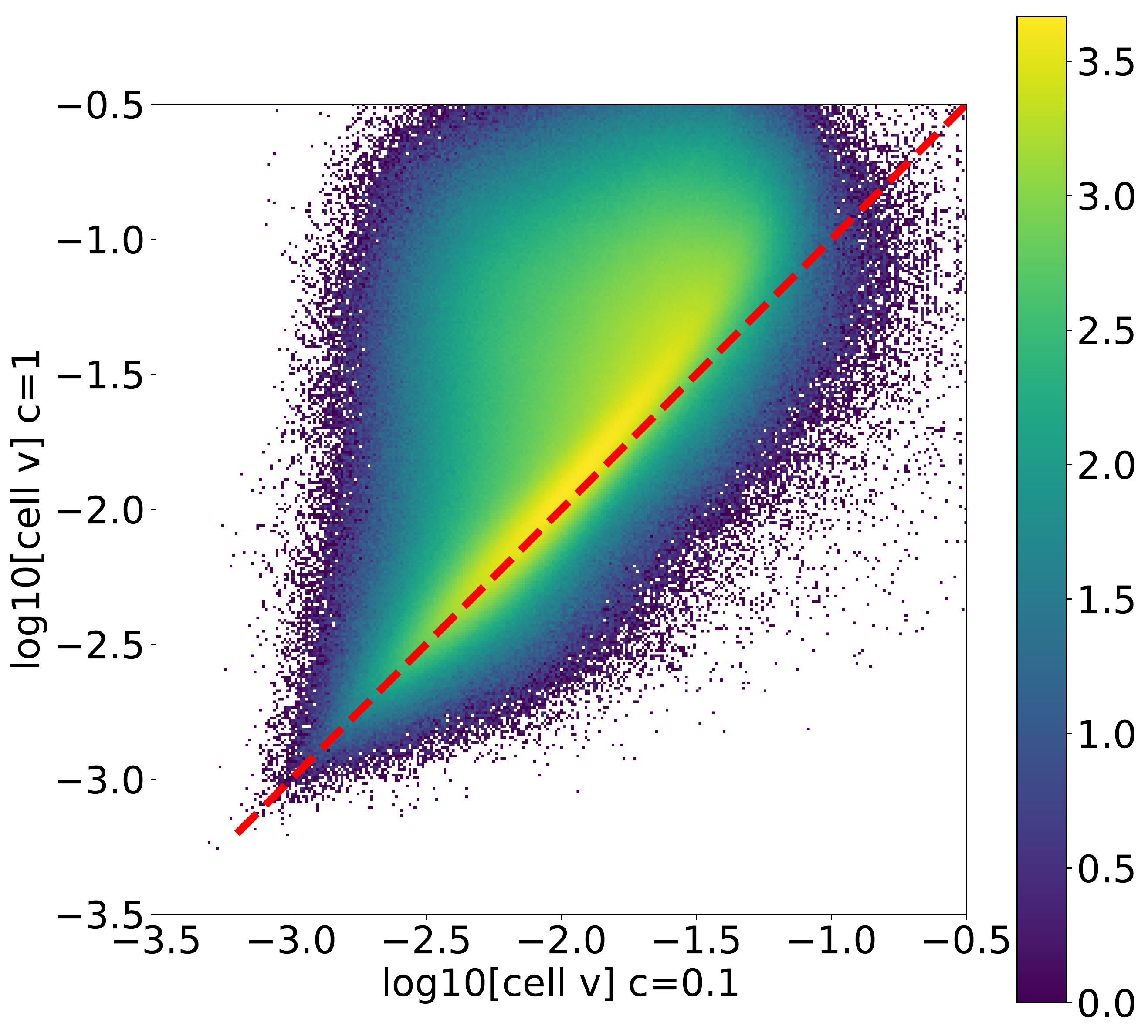}
\caption{{Cell-by-cell comparison of reionization times (top) and front velocities (bottom) in the (64 Mpc/h)$^3$ simulations with $\tilde c=1.0$ and $\tilde c=0.1$. The logarithm of cell number counts in each bins of reionization times/front velocities are color-coded, while the dashed lines stand for the 1:1 correspondence. }}
\label{fig:comp_cell}
\end{figure}

\end{appendix}
\end{document}